\documentclass[preprint,number,12pt]{elsarticle}
\usepackage{threeparttable}  % Revisions added by Zunfeng Wang for table adding Apr. 29, 2011

\usepackage{amssymb}
\makeatletter

\newcommand{\Rmnum}[1]{\expandafter\@slowromancap\romannumeral #1@}

\makeatother
\begin{document}
\begin{frontmatter}
\title{Cycles of strategies and changes of distribution in public goods game: An experimental investigation}
\author{Bin Xu$^{a,b}$\footnote{Corresponding author. Tel.: +86 15888897050; fax: +86 571 28008333.
E-mail addresses: xubin211@zju.edu.cn (B. Xu).}}

\address{$^{a}$Public Administration College, Zhejiang Gongshang University, Hangzhou, 310018, China}
\address{$^{b}$Experimental Social Science Laboratory, Zhejiang University, Hangzhou, 310058, China}

\begin{abstract}
In this communication, a simple mechanism in the optional public goods game is experimentally investigated using two experimental settings; and first time, the cyclic strategy pattern in full state space is demonstrated by means of velocity. It is, furthermore, elaborated that the strategies of cooperation, defection and nonparticipant form a Rock-Paper-Scissors type cycle, and the cycle of three strategies are persistent over 200 rounds. This cycle is very similar to the cycle given by evolutionary dynamics e.g. replicator dynamics. The mechanism that nonparticipant can sustain cooperation is driven by the Rock-Paper-Scissors type of cyclic dominance in the three strategies. That is, if the cycle is existent, the cooperation will always sustain. Meanwhile, the distribution of social states changes in the state space and from cooperation as the most frequent strategy to defection and, from defection to nonparticipant, forms a clear rotation path in a long run. These results seem to implicate that the evolutionary dynamics has ability to capture the real dynamics applying not only on biosphere, but also on human society.
\end{abstract}




%\title{}


%\author{}
%
%\address{}
%
%\begin{abstract}
%%% Text of abstract
%
%\end{abstract}
%
%\begin{keyword}
%%% keywords here, in the form: keyword \sep keyword
%
%%% MSC codes here, in the form: \MSC code \sep code
%%% or \MSC[2008] code \sep code (2000 is the default)
%
%\end{keyword}

 \end{frontmatter}

% \linenumbers

%\section{}
%\label{}

\tableofcontents
\newpage

\section{Introduction}

Dissolving the "tragedy of the commons"~\cite{garrett1968tragedy} is a central concern of studying cooperation in humans. Numerous studies have been conducted applying the "public goods" game to study the problem of maintaining cooperation in a group of unrelated individuals. The society achieves its highest payoff if every individual contributes; however, defector always gets more than cooperator. In typical examples, the individual contributions are multiplied by a factor $r$ and then divided equally among all players. Where $r$ is smaller than the group size, which is an example of a social dilemma~\cite{schelling1973hockey,dawes1980social}. Although most of the people do not contribute zero to a common pool in the one-shot public goods game or in the initial stage of repeated public goods game, but the contributions usually decline over a few rounds~\cite{ledyard1995public}. Vast experimental work on public goods game has focused on two different conditions which would lead to high contributions to the public pool, punishment~\cite{yamagishi1986provision,fehr2002altruistic,sigmund2007punish,
masclet2003monetary,casari2005design,anderson2006non,
carpenter2007demand,nikiforakis2008comparative,gachter2008long,dreber2008winners,milinski2008human,
herrmann2008antisocial,henrich2006costly,gachter2009reciprocity,gurerk2006competitive} or/and reward~\cite{sefton2007effect,sutter2010choosing,dickinson2001carrot,andreoni2003carrot,
fehr1997reciprocity,rockenbach2006efficient,rand2009positive}, and reputation~\cite{nowak1998evolution,
wedekind2000cooperation,bolton2005cooperation,seinen2006social,milinski2002reputation,
barclay2004trustworthiness,panchanathan2004indirect,semmann2004strategic,semmann2005reputation,
milinski2006stabilizing,shinada2008bringing}. However, Christoph Hauert et al.~\cite{hauert2002volunteering,hauertwz2002replicator} pointed out, reciprocal altruism fails to provide a solution if interactions are not repeated often enough or groups are too large. Punishment and reward can be very effective but require that defectors must be traced and identified.

In 1993, John M. Orbell and Robyn M. Dawes~\cite{orbell1993social} outlined a model of freedom to choose between playing and not playing particular Prisoner's Dilemma games and conducted experiments which showed that: (1) Social welfare and the relative welfare of intending cooperators are higher when subjects are free to choose between entering and not entering particular Prisoner's Dilemma relationships; and this difference is a consequence of intending cooperator's greater willingness to enter such relationship, not because of any capacity to recognize and avoid intending defectors. In 1995, John Batali and Philip Kitcher~\cite{batali1995evolution} also compared the evolution of altruism in optional and compulsory games. In their theoretical analysis, populations playing the compulsory game can become stuck in states of low cooperation that last many generations, while the optional games provide routes out of such states to states of high cooperation. And their computational simulations of the evolution of populations playing these games support their analytic results.

Christoph Hauert et al.~\cite{hauert2002volunteering,hauertwz2002replicator} presented models introducing a third option named non-participation into public goods game. That means, people can either join the public goods game and then choose cooperate (C), defect (D), or refuse to participate in the game (N). The game are denoted, hereafter, as NCD games. Then a cycle of the three strategies, Rock-Paper-Scissors type, is predicted. When cooperators are more frequent, defectors can exploit a large group of cooperators, whereas loners have the highest profit when defectors are frequent. When loners are most frequent, the public group size is reduced inviting cooperation because the game is no longer a dilemma in small groups~\cite{schelling1973hockey,dawes1980social,boyd1988evolution,sober1999unto}. Hence, volunteering relaxes the social dilemma: instead of defectors winning the world, coexistence among cooperators, defectors and loners is expected\cite{michor2002evolution}. Later, Tatsuya Sasaki et al.~\cite{sasaki2012take} investigated punishment and reward effect in this public goods game. Thus, nonparticipant as a natural mechanism is theoretically well studied, and the cyclic pattern in the full state space is clearly exhibited.

Dirk Semmann et al.~\cite{semmann2003volunteering}, following the above theoretical approach, conducted an experiment in which a sample group of 6 players was randomly chosen from a large population of 14. They manipulated initial conditions and produced each predicted direction. By manipulating displayed decisions, it is pretended that if the defectors have the highest frequency, the loners soon become most frequent, as do the cooperators after the loners and defecators. On average, cooperation is perpetuated at a substantial level. However, the results, based on the initial conditions they manipulated, didn't come out in the subsequent real experiment over 50 rounds. It is not clear, whether the Rock-Paper-Scissors type strategy cycle exists in the real experiment or not. That means, in the perspective of evolution, the empirical cyclic dynamic pattern in the full state space has not been proved experimentally.

In this article, the mechanism in the optional public goods game is investigated experimentally, and first time, the cyclic strategy pattern in full state space is demonstrated by means of velocity. The remaining of this article is organized as follows: section two describes the experimental design and procedure; section three demonstrates the results; and the last discusses and concludes the results.

\section{Experimental design and procedure}
\subsection{Game and payoffs}

The public goods game, which is employed here, is actually known as optional public goods game~\cite{hauert2002volunteering,hauertwz2002replicator}. Different with compulsory public goods game, there are three options, viz. cooperate (C), defect (D) and refuse to participate in the game (N), which are denoted, as NCD games. That means, the subject not willing to participate in the public goods game can choose N, but if he participates, he has two options, C and D. If only one person participates in the game, then, we think he will be equivalent to a loner. The public goods game starts only when two or more persons are there to participate in.

The parameters used in our experiments are similar to that Tatsuya Sasaki et al. have used in their theoretical analysis~\cite{sasaki2012take}, where nonparticipation with intermittent cooperation bursts and form a Rock-Paper-Scissors type cycle, N-C-D-N. From time to time, $n=5$ players are faced with an opportunity to participate in a public goods game. If the subject refuses to participate, and chooses N, he gets 0, but if he decides to participate, he pays at a cost $g=0.5$ irrespective whether he chooses C or D. If the subject chooses C, then he will contribute a fixed amount $c=1$, which will be be multiplied by $r=3$ and will distributed equally among all other participators of the group (except himself). If the subject chooses D, he will contribute 0 and share the cooperator's contribution. The payoffs for cooperator, defector and Nonparticipator are $P_{c}$, $P_{d}$, and $P_{n}$, respectively, and is given by:

\begin{eqnarray}
P_{c}=-g-c+rc\frac{n_{c}-1}{n_{c}+n_{d}-1},\\
P_{d}=-g+rc\frac{n_{c}}{n_{c}+n_{d}-1},\\
P_{n}=0,
\label{eq:payoff}
\end{eqnarray}
where $n_{c}$ is the number of cooperators and $n_{d}$ is the number of defectors. Therefore, if all subjects choose C, then each of them gets 1.5; if all subjects choose D, then each of them gets -0.5; the defector will always get additional ($c+\frac{rc}{n_{c}+n_{d}-1}$) than the cooperator only if there is a cooperator. So, social dilemma appears since the group will get the maximum payoff if all subjects choose C, however defector always gets more than cooperator. The nonparticipator will get 0 at all conditions. Therefore, the option N provides an opportunity to escape the deadlock of defection. When N is frequent, cooperation is invited. Again, once cooperator emerges, the defector will take advantage of cooperator. The Rock-Paper-Scissors type strategy cycles are expected.

\subsection{Population size}

According to the evolutionary dynamics~\cite{hauert2002volunteering,hauertwz2002replicator,sasaki2012take}, from time to time, sample group of N players are randomly chosen from a large population and asked to participate in a single public goods game. Following the theoretical research~\cite{hauert2002volunteering,hauertwz2002replicator,sasaki2012take} and the experimental investigation~\cite{semmann2003volunteering}, we set a treatment in which, from time to time, a sample group of 5 players is randomly chosen from a large population of 13. Besides this, we add another treatment of a population size of 5, that means the five players are fixed in a group and interact with each other from time to time. This setting allows us to investigate the population itself instead of sample group, and to trace the individual's behavior period by period. We call fixed group T1 and sample group T2.

\subsection{Information feedback}

For treatment 1, where the group is fixed, the feedback includes every player's strategy used in the previous round and the corresponding score. Particularly, every player has a unique and unaltered number in the group. The feedback for treatment 2 includes the frequency of strategy C, D and N, used in the sample group of 5 players and the corresponding payoff of every strategy. And this information is provided not only to members of the sample group but also to unchosen members remained in the population in the previous round.

In summary, the two treatments are exhibited in Table~\ref{tab:Treatment summarize}.

\begin{table}[htbp2]
\centering
\begin{threeparttable}
\small
\caption{\label{tab:Treatment summarize} Treatment Summary}
\begin{tabular}{cccc}
  \hline
   \hline
Treatment	&Population Size &Feedback&Groups		\\
   \hline
T1			&5&every player's strategy and score&12		\\
T2		&13&the number of strategy and relative score	&6		\\

  \hline
 \hline
  \end{tabular}
     \end{threeparttable}
  \end{table}

\subsection{Experimental procedure}

The experiments were conducted in the Experimental Social Science Laboratory of Zhejiang University from 27 May, 2012 to 26 April, 2013. Each session lasted about 1 hour including the introduction and a quiz via computer. Each subject was made to sit on an isolated seat with a computer, no communication was allowed during the experiment. The software for the experiments was developed as a web based system designed by the authors. Every subject made decision by using 'C', 'D' or 'N' button. After the last one submitted the decision, every player received the feedback. Each session consisted 200 rounds. At the conclusion of the experiments, every subject obtained the cumulative score (CS). Each subject's CS ($CS\times13/5$, in T2) including initial account (225) was converted into RMB currency according to $0.1\times CS$ as payments. In addition, each participator was paid 5 Yuan RMB as the showing up fee. The average payoff was 28 Yuan RMB. All 138 $(5\times12+13\times6)$ students of Zhejiang Univerisity majoring in different areas were recruited into these experiments and each subject participated only in one session.

\section{Results}

\subsection{The distribution of strategy use}
The average cooperation rate of group are $0.224$ and $0.096$, in T1 and T2, respectively, while the defection rate of group are $0.385$ and $0.466$, respectively. We are also interested in the cooperation rate and defection rate among the volunteers, i.e. given the condition of participation, what would be the cooperation rate and defection rate? We call them the conditional cooperation and defection rate. The conditional cooperation rates are 0.363 in T1 and 0.169 in T2, while the conditional defection rates are 0.637 in T1 and 0.831 in T2, see Table~\ref{tab:strategyusesummarize}.

\begin{table}[htbp2]
\centering
\begin{threeparttable}
\small
\caption{\label{tab:strategyusesummarize} Strategy Use of Group}
\begin{tabular}{c|lccccc}
  \hline
   \hline
Treatment	&Variable	&Obs	&Mean	&Std.Dev.	&Min	&Max	\\
\hline
T1	&cooperation rate	            &12	&0.224	&0.085	&0.102	&0.340	\\
T1	&defection rate	                &12	&0.385	&0.092	&0.230	&0.527	\\
T1	&nonparticipant rate	        &12	&0.391	&0.141	&0.162	&0.609	\\
T1	&conditional cooperation rate	&12	&0.363	&0.092	&0.225	&0.493	\\
T1	&conditional defection rate	    &12	&0.637	&0.092	&0.507	&0.775	\\
   \hline
T2	&cooperation rate	            &6	&0.096	&0.033	&0.072	&0.151	\\
T2	&defection rate	                &6	&0.466	&0.068	&0.367	&0.552	\\
T2	&nonparticipant rate	        &6	&0.438	&0.093	&0.324	&0.560	\\
T2	&conditional cooperation rate	&6	&0.169	&0.036	&0.124	&0.229	\\
T2	&conditional defection rate	    &6	&0.831	&0.035	&0.771	&0.876	\\
    \hline
 \hline
  \end{tabular}
     \end{threeparttable}
  \end{table}

The cooperation rates in both treatments are larger than 0 ($p$=0.0000, $n_1$=12 in T1 and $p$=0.0004, $n_2$=6 in T2, $t$-test). The cooperation rate of T1 is larger than that of T2 ($p$=0.002, $z$=3.092, $n_1$=12 and $n_2$=6, Mann-Whitney test), the conditional cooperation rate of T1 is also larger than that of T2 ($p$=0.0015, $z$=3.184, $n_1$=12 and $n_2$=6, Mann-Whitney test), while the defection rate of T1 is smaller than that of T2 ($p$=0.075, $z$=-1.780, $n_1$=12 and $n_2$=6, Mann-Whitney test), again, the conditional defection rate of T1 is smaller than that of T2 ($p$=0.0015, $z$=-3.184, $n_1$=12 and $n_2$=6, Mann-Whitney test). Interestingly, there is no significant difference between two treatments on nonparticipant rate ($p$=0.5741, $z$=-0.562, $n_1$=12 and $n_2$=6, Mann-Whitney test).

\subsection{The overall trend of strategy use over time}

\begin{figure}
\centering
\includegraphics[angle=0,width=6cm]{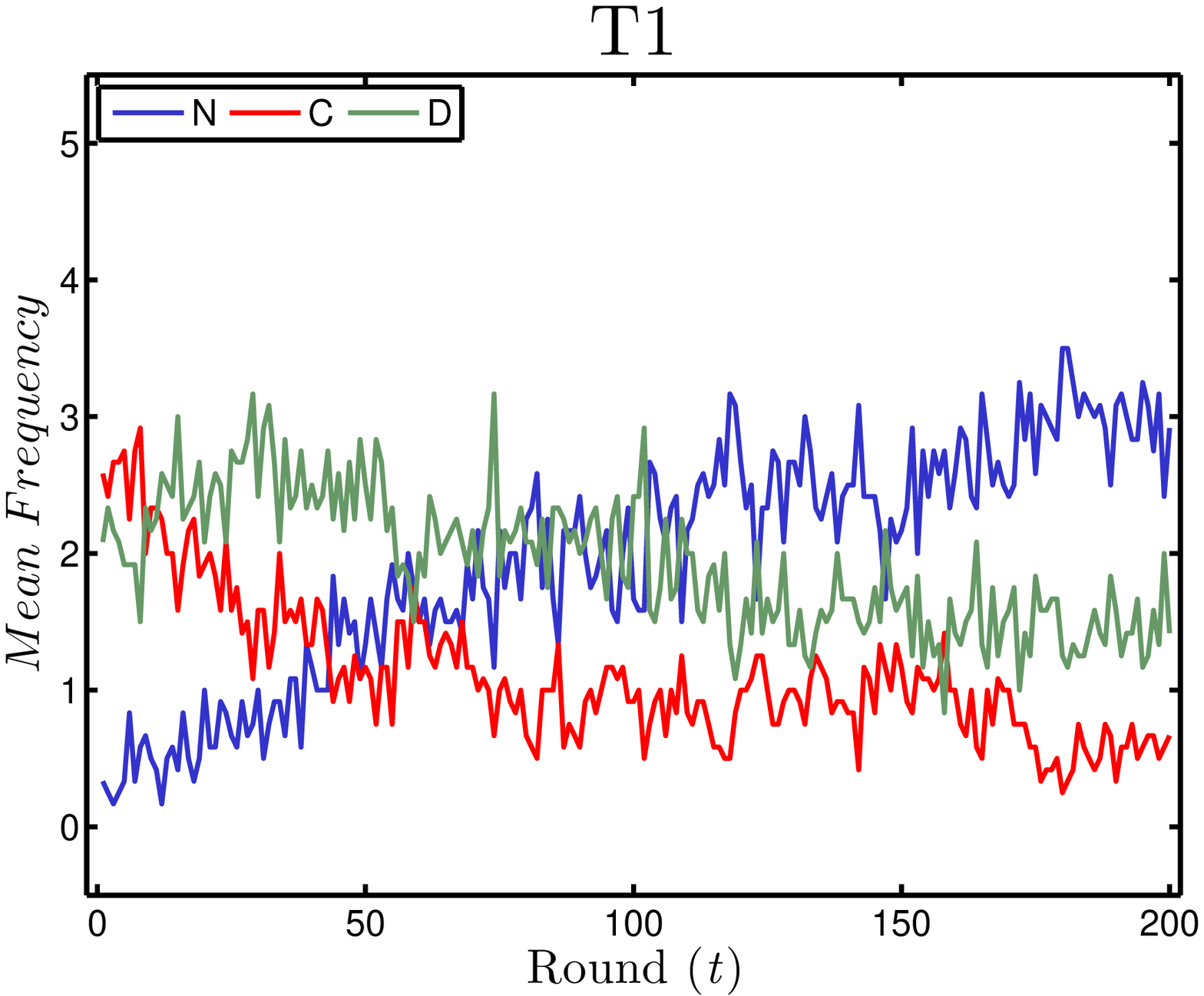}
\includegraphics[angle=0,width=6cm]{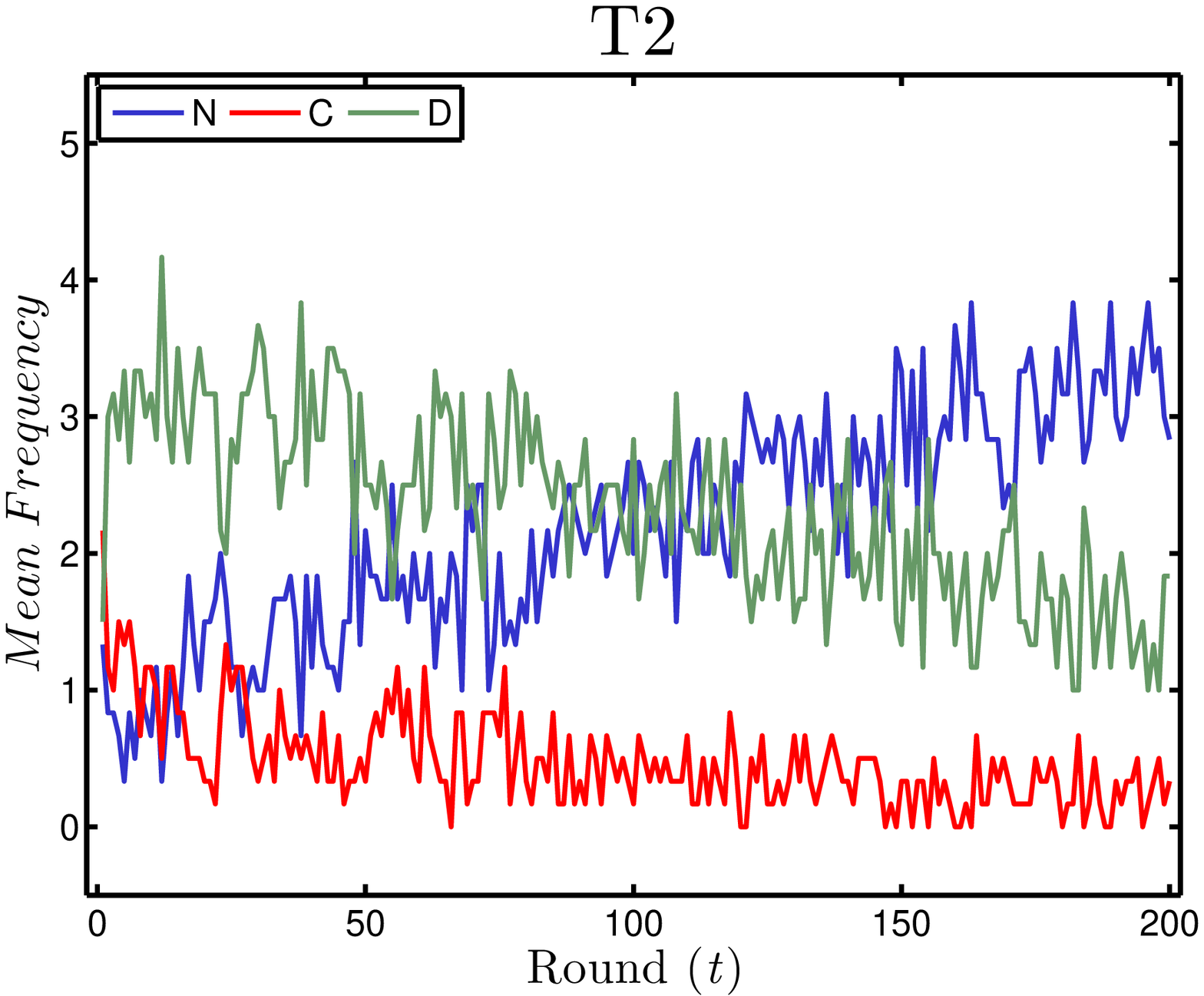}
\caption{The strategy use over time. The left panel is T1 and the right one is T2.\
\label{fig:strategy ues over time}}
\end{figure}

The strategy use keeps changing over 200 rounds. At the beginning of the game, the strategy C is used frequently, and the strategy N is used rarely. The cooperation rate of first round is $0.52\pm0.20$ (mean $\pm$s.d.) in T1 and $0.43\pm0.23$ (mean $\pm$s.d.) in T2, respectively. The nonparticipant rate of first round is $0.07\pm0.13$ (mean $\pm$s.d.) in T1 and $0.27\pm0.21$ (mean $\pm$s.d.) in T2. The defection rate of first round is $0.41\pm0.20$ (mean $\pm$s.d.) in T1 and  $0.30\pm0.11$ (mean $\pm$s.d.) in T2. However, the cooperation rate declines quite quickly while the strategies D and strategy N increase. Later, the strategy D is used over strategy C, and it becomes the most frequent strategy, and the defection rate comes to it's pick value, $0.63\pm0.24$ (mean $\pm$s.d.) at round 29 in T1 and $0.83\pm0.23$ (mean $\pm$s.d.) at round 12, in T2. After that, the defection rate keeps dropping with fluctuations, but remains over the cooperation rate. After half of the experimental process, the strategy N become the most frequently used strategy, and this result is maintained till end. At the last round, the cooperation rate is $0.13\pm0.16$ (mean $\pm$s.d.) in T1 and $0.07\pm0.10$ (mean $\pm$s.d) in T2; the defection rate is $0.28\pm0.26$ (mean $\pm$s.d.) in T1 and $0.37\pm0.29$ (mean $\pm$s.d) in T2; the nonparticipant rate is $0.58\pm0.35$ (mean $\pm$s.d.) in T1 and $0.57\pm0.32$ (mean $\pm$s.d) in T2. The change of strategy use over time is illustrated in Figure~\ref{fig:strategy ues over time}.

It seems that there is a sequence in strategy use, i.e. C--D--N in above analysis. However, after N, the strategy C does not remain the most frequent strategy, thus makes the Rock-Paper-Scissors type cycle of strategy imperfect.

\begin{figure}
\centering
\includegraphics[angle=0,width=2cm]{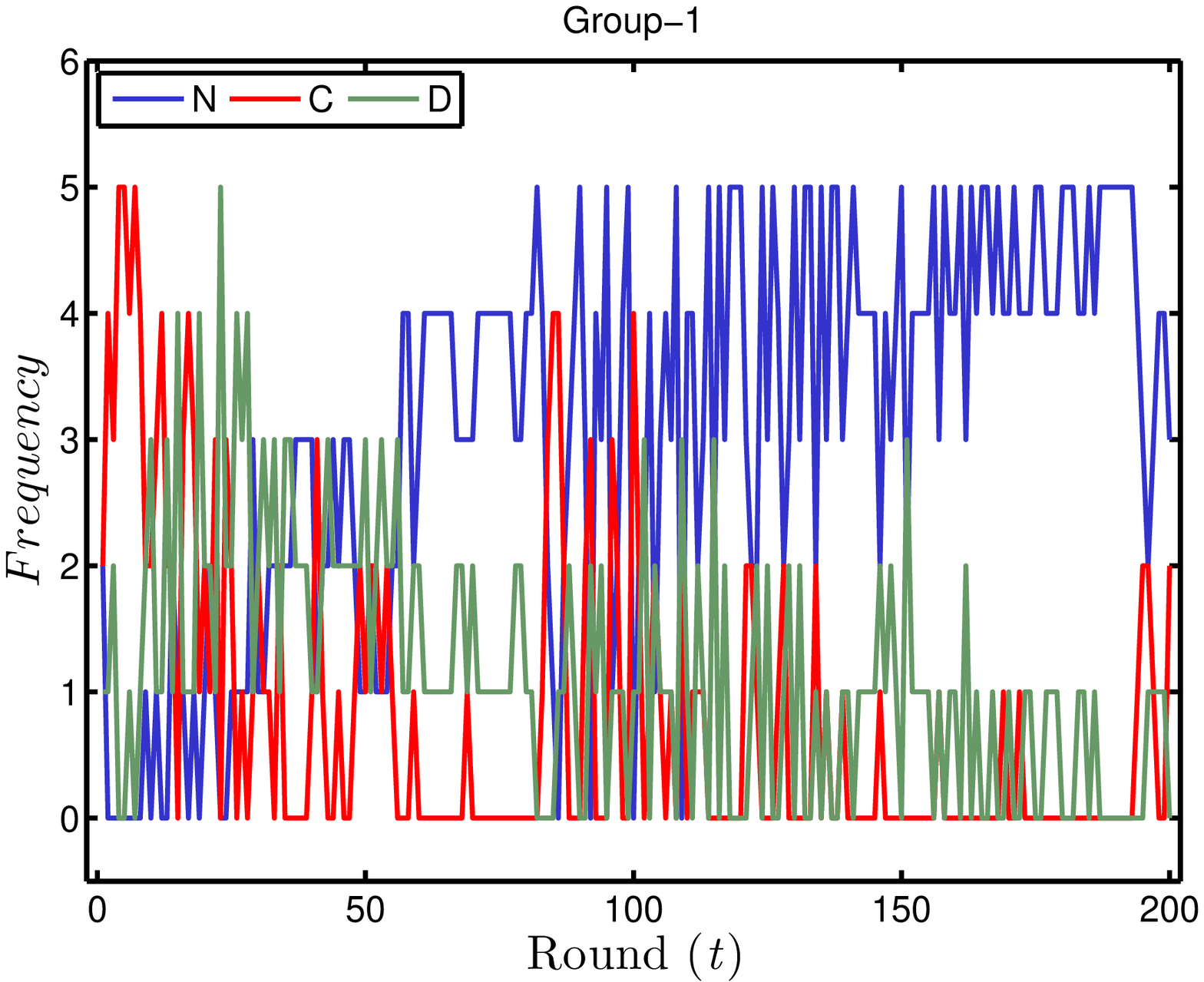}
\includegraphics[angle=0,width=2cm]{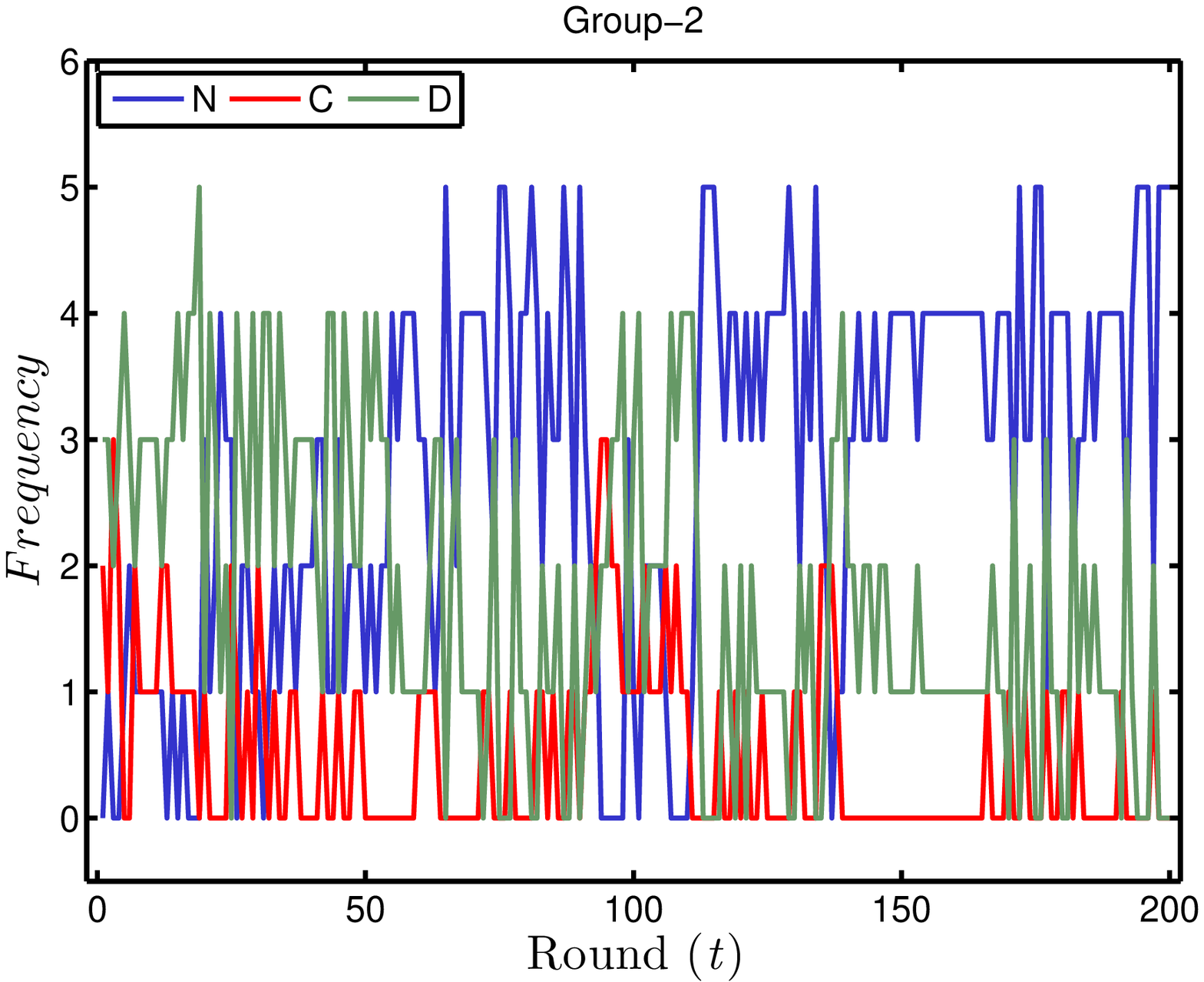}
\includegraphics[angle=0,width=2cm]{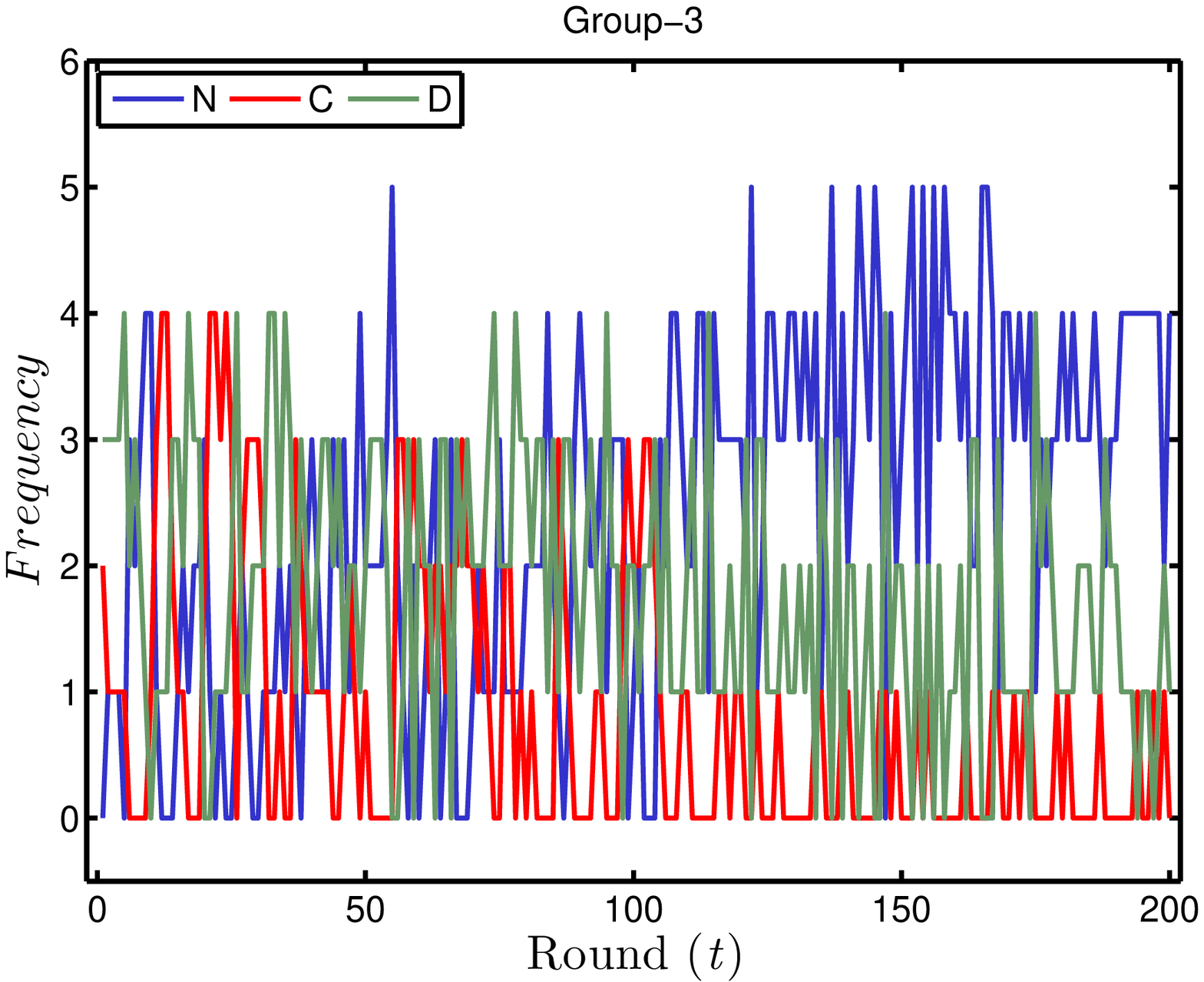}
\includegraphics[angle=0,width=2cm]{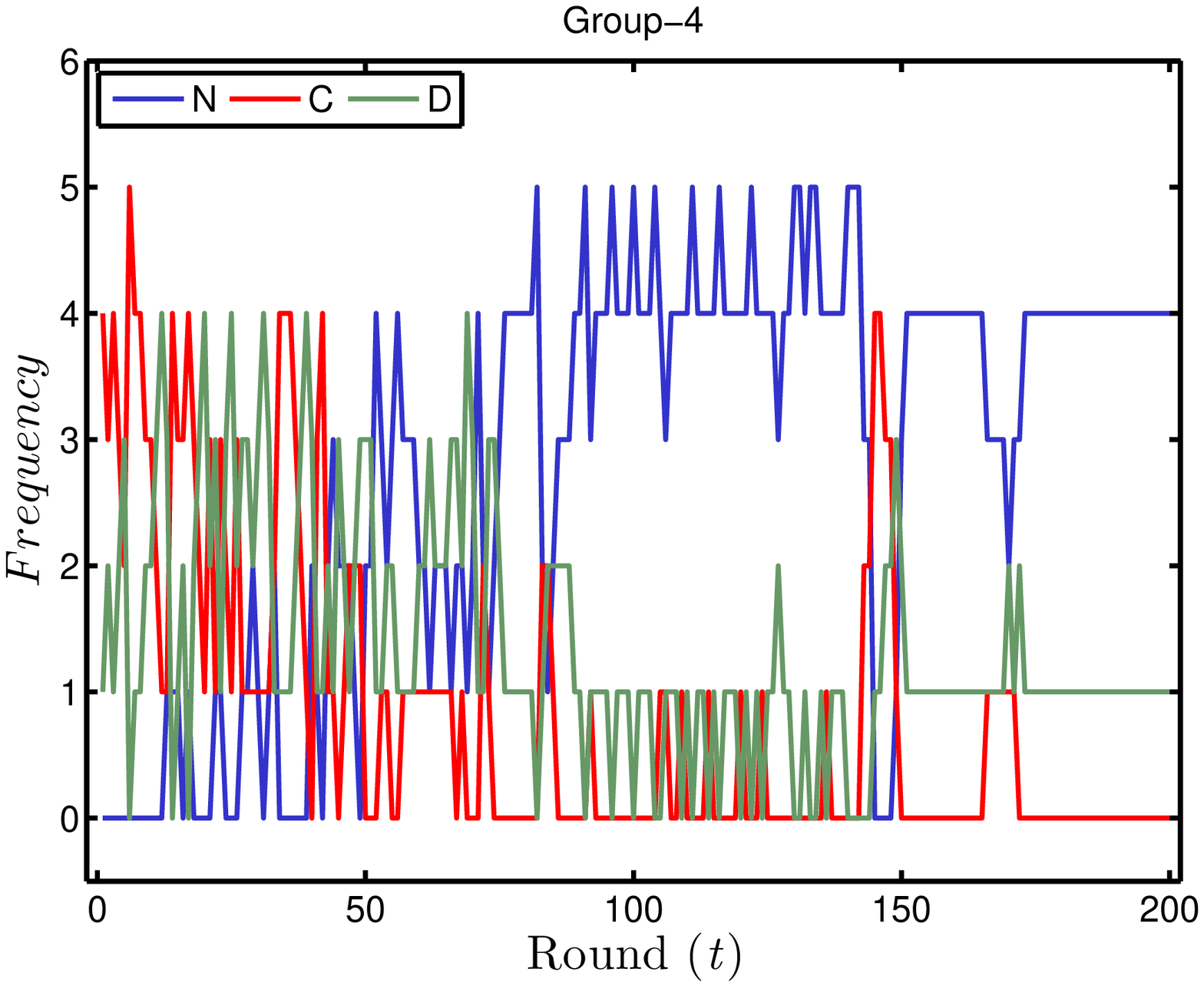}
\includegraphics[angle=0,width=2cm]{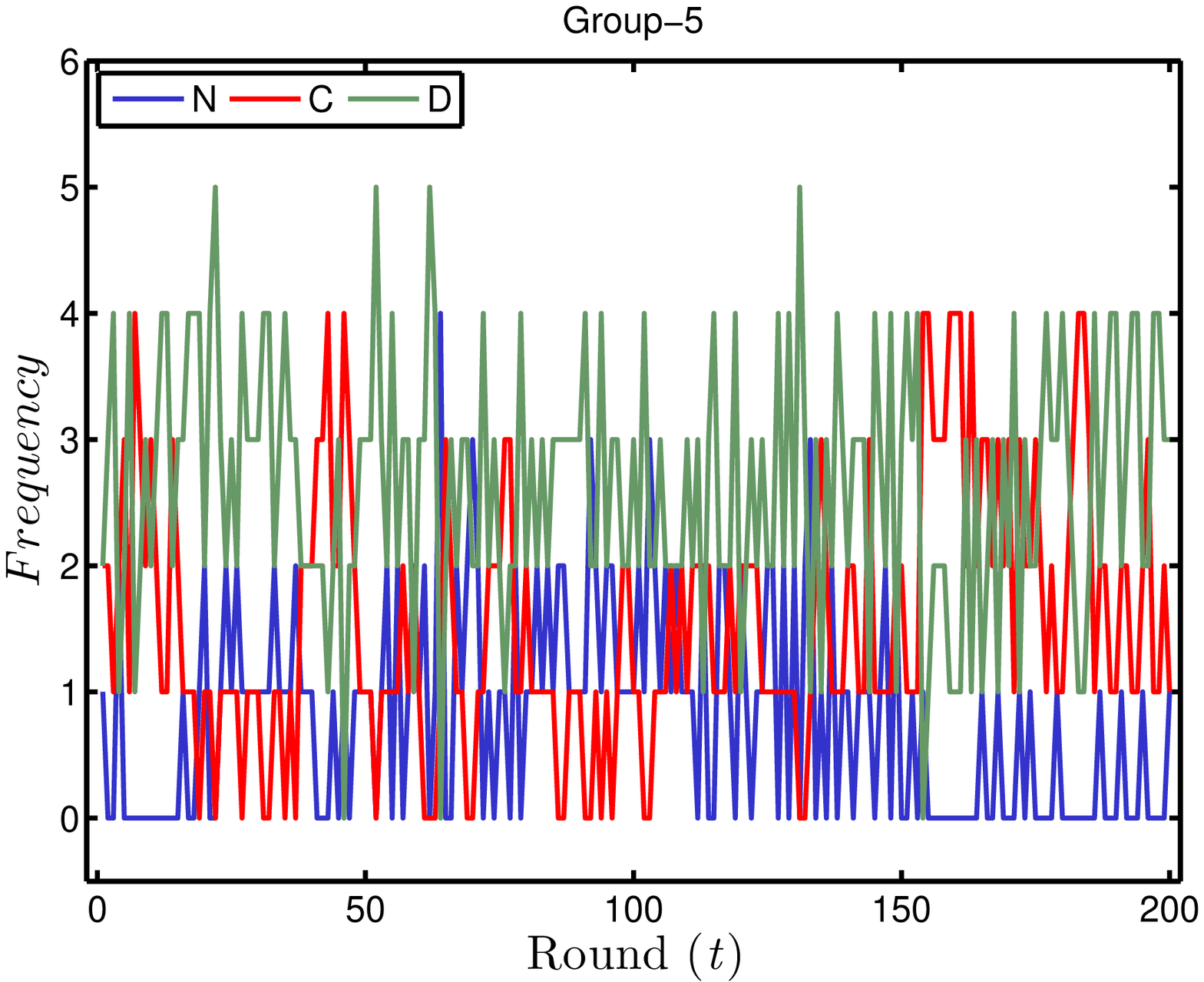}
\includegraphics[angle=0,width=2cm]{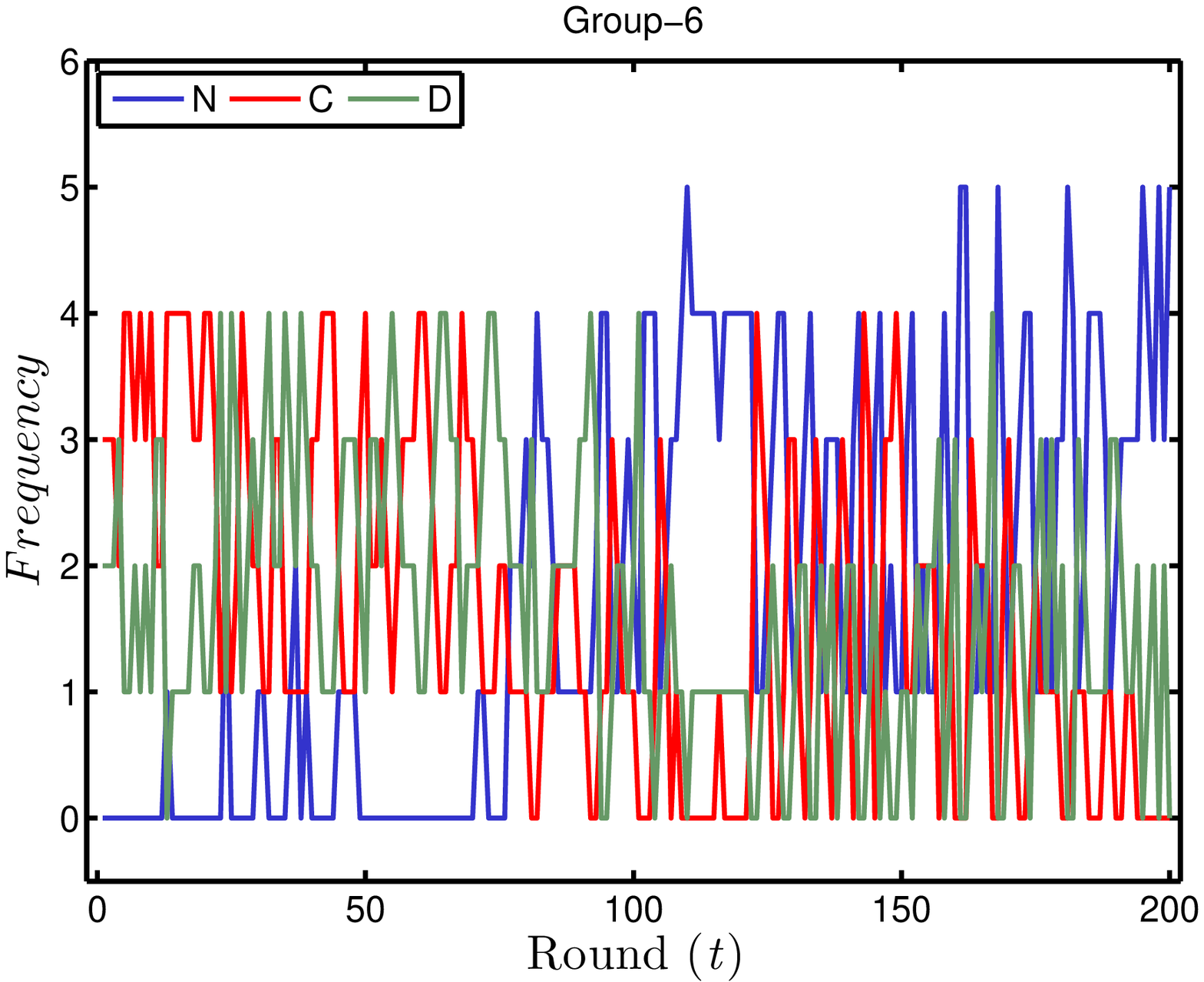}
\includegraphics[angle=0,width=2cm]{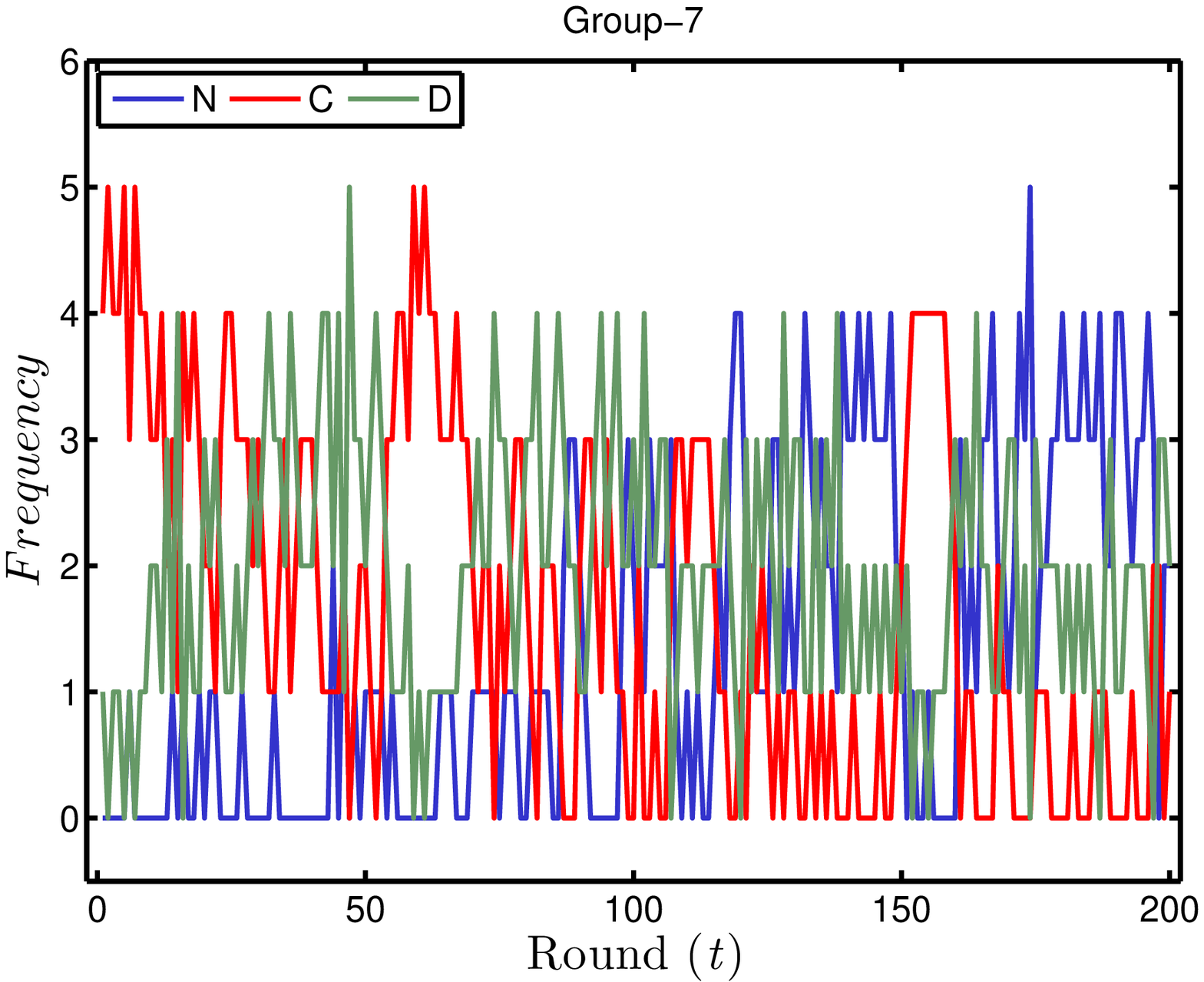}
\includegraphics[angle=0,width=2cm]{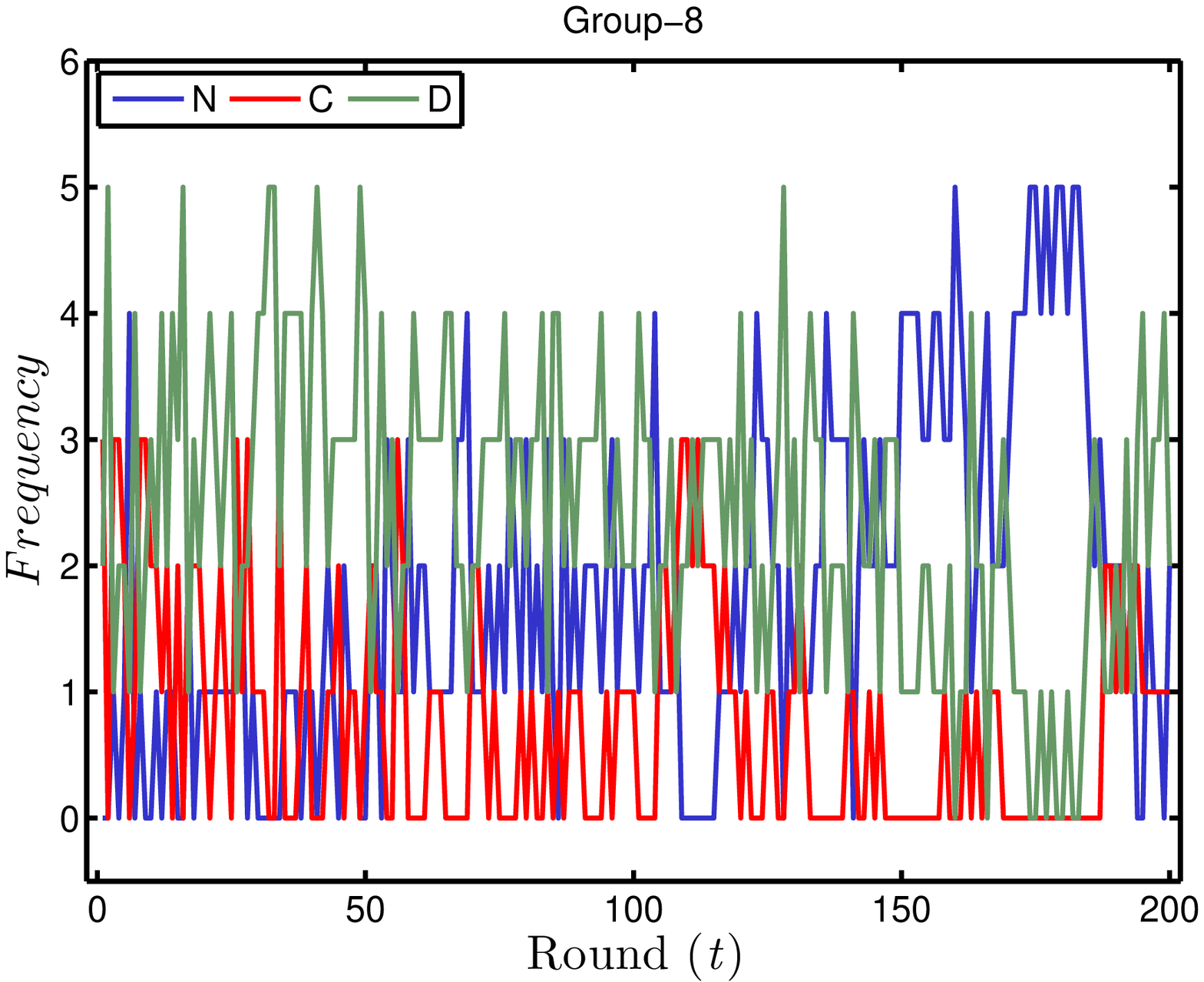}
\includegraphics[angle=0,width=2cm]{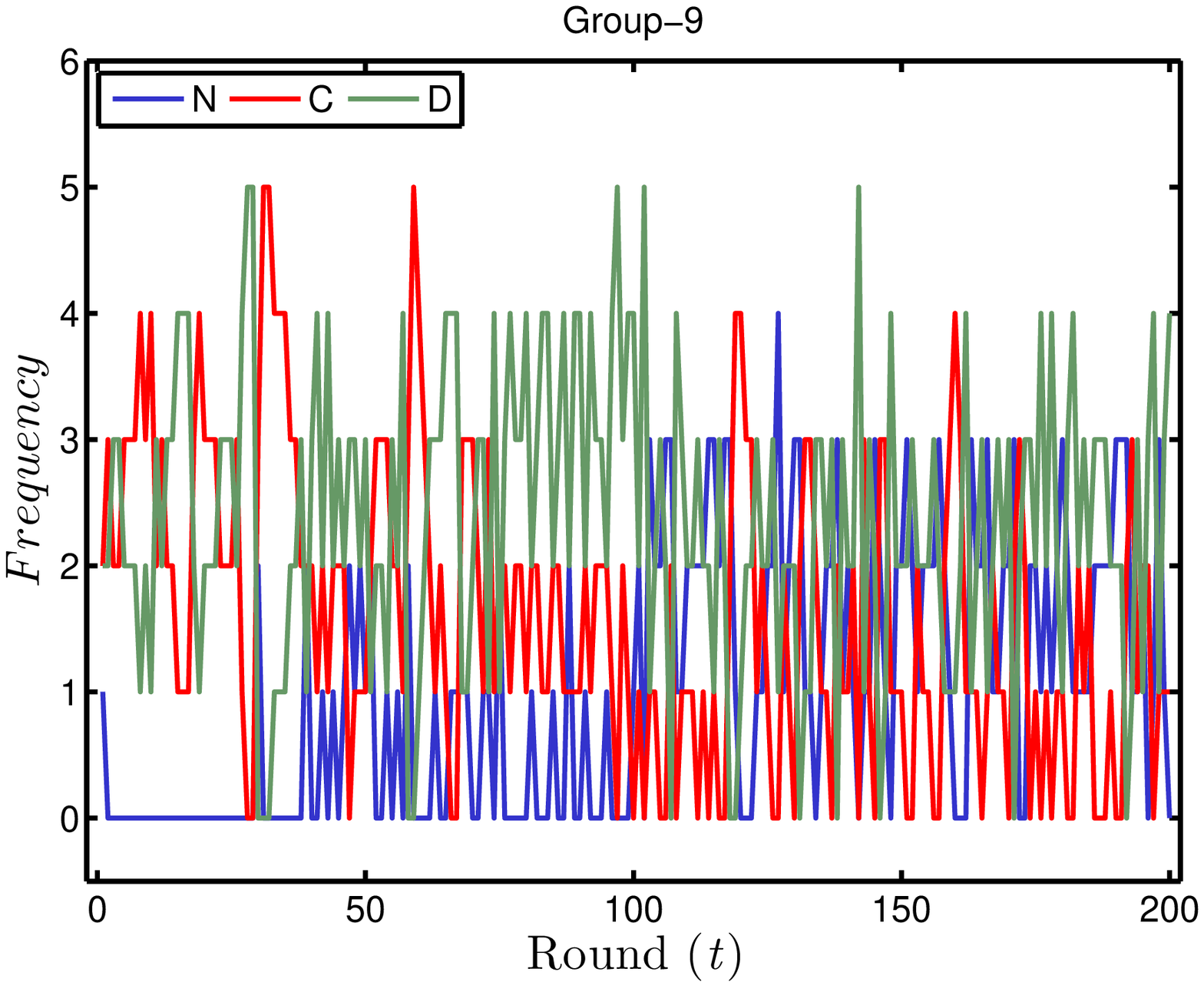}
\includegraphics[angle=0,width=2cm]{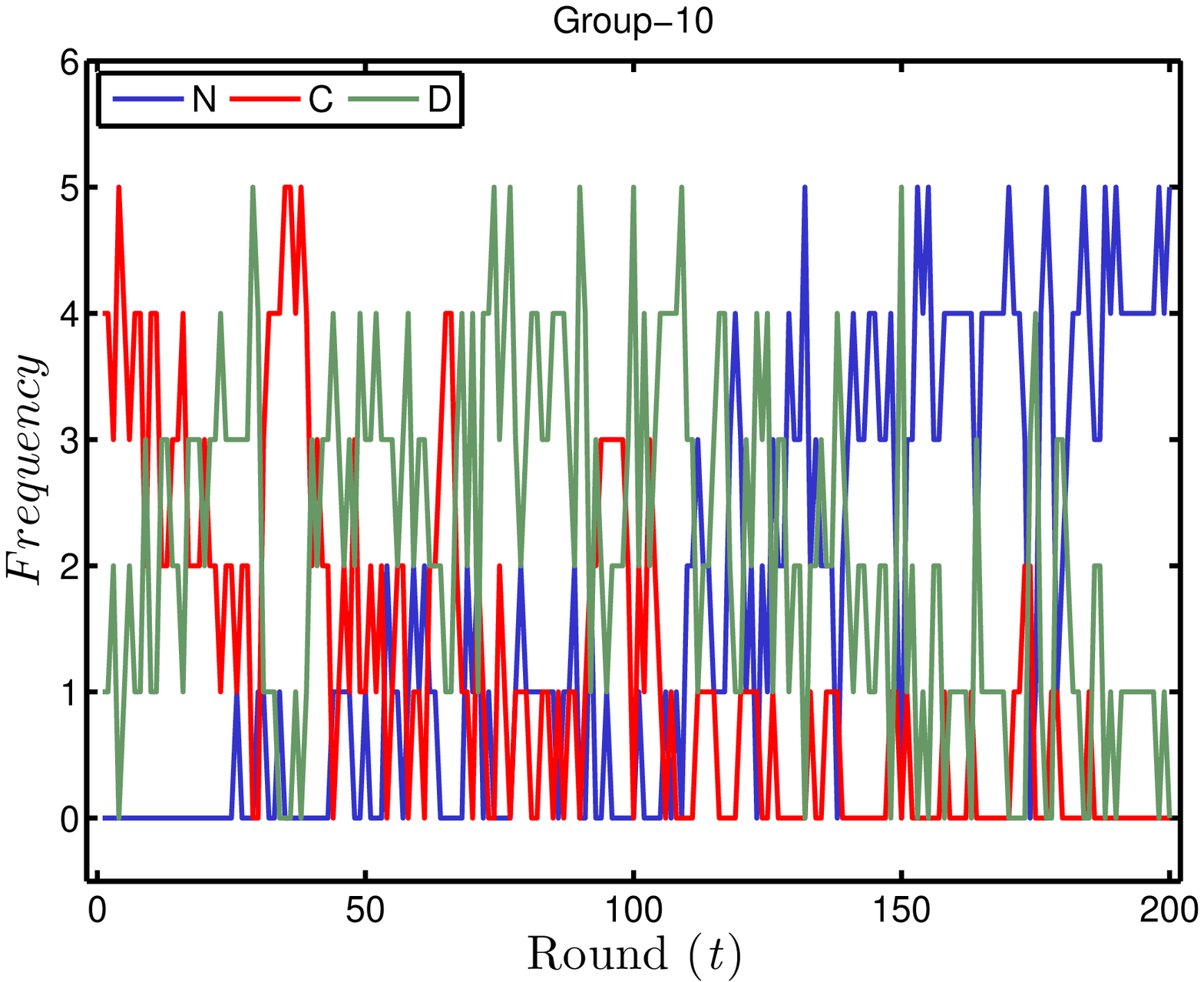}
\includegraphics[angle=0,width=2cm]{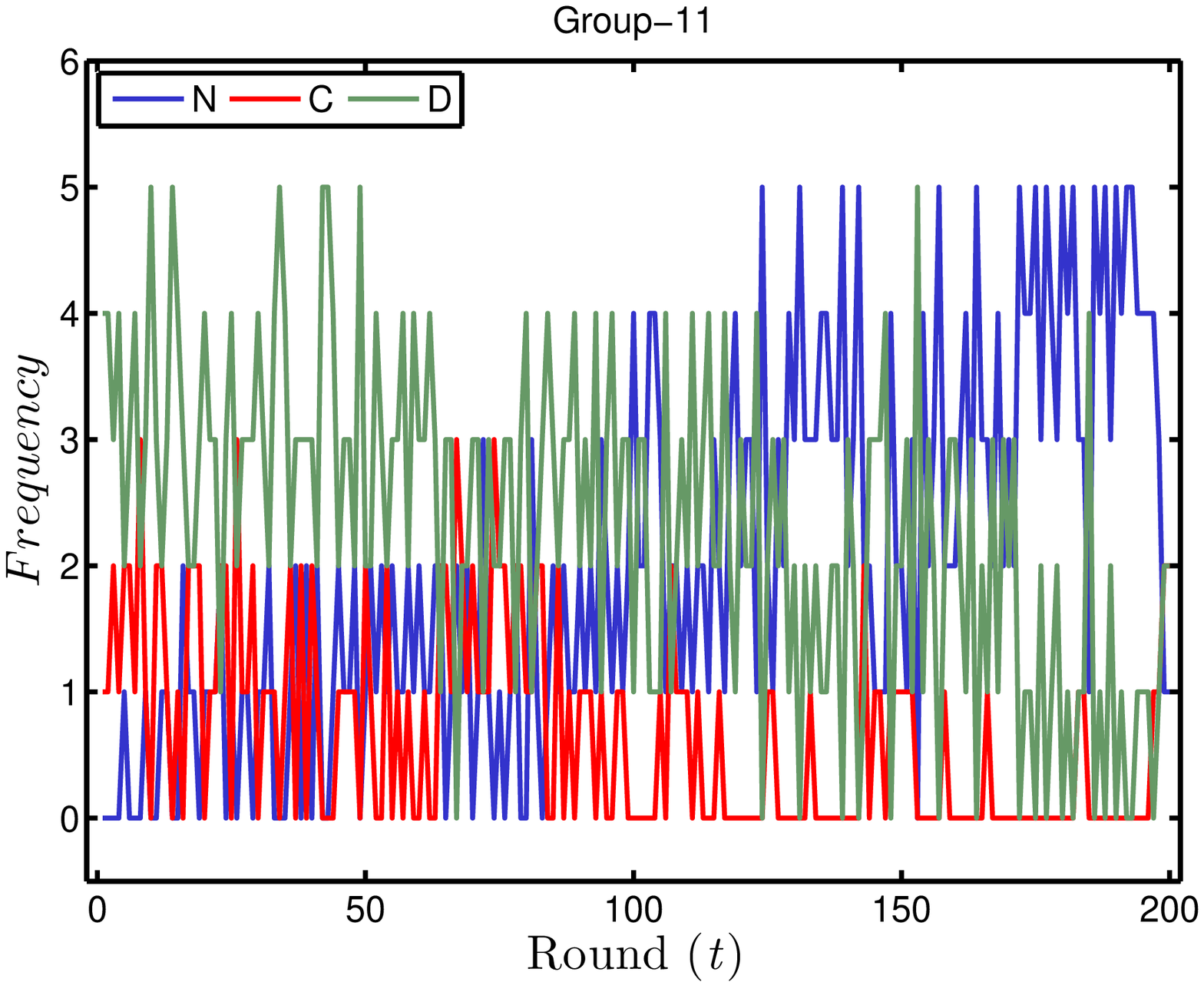}
\includegraphics[angle=0,width=2cm]{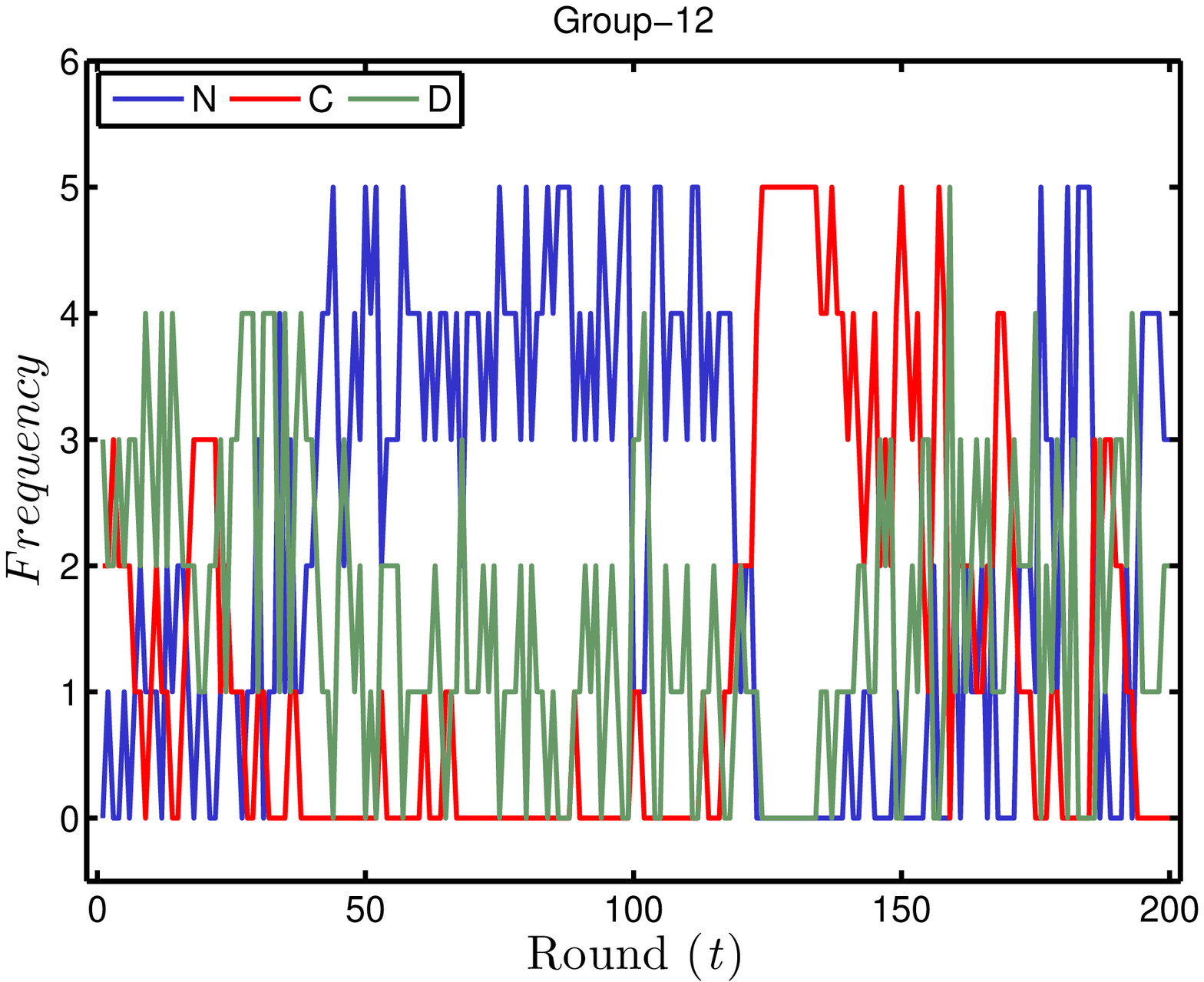}
\includegraphics[angle=0,width=2cm]{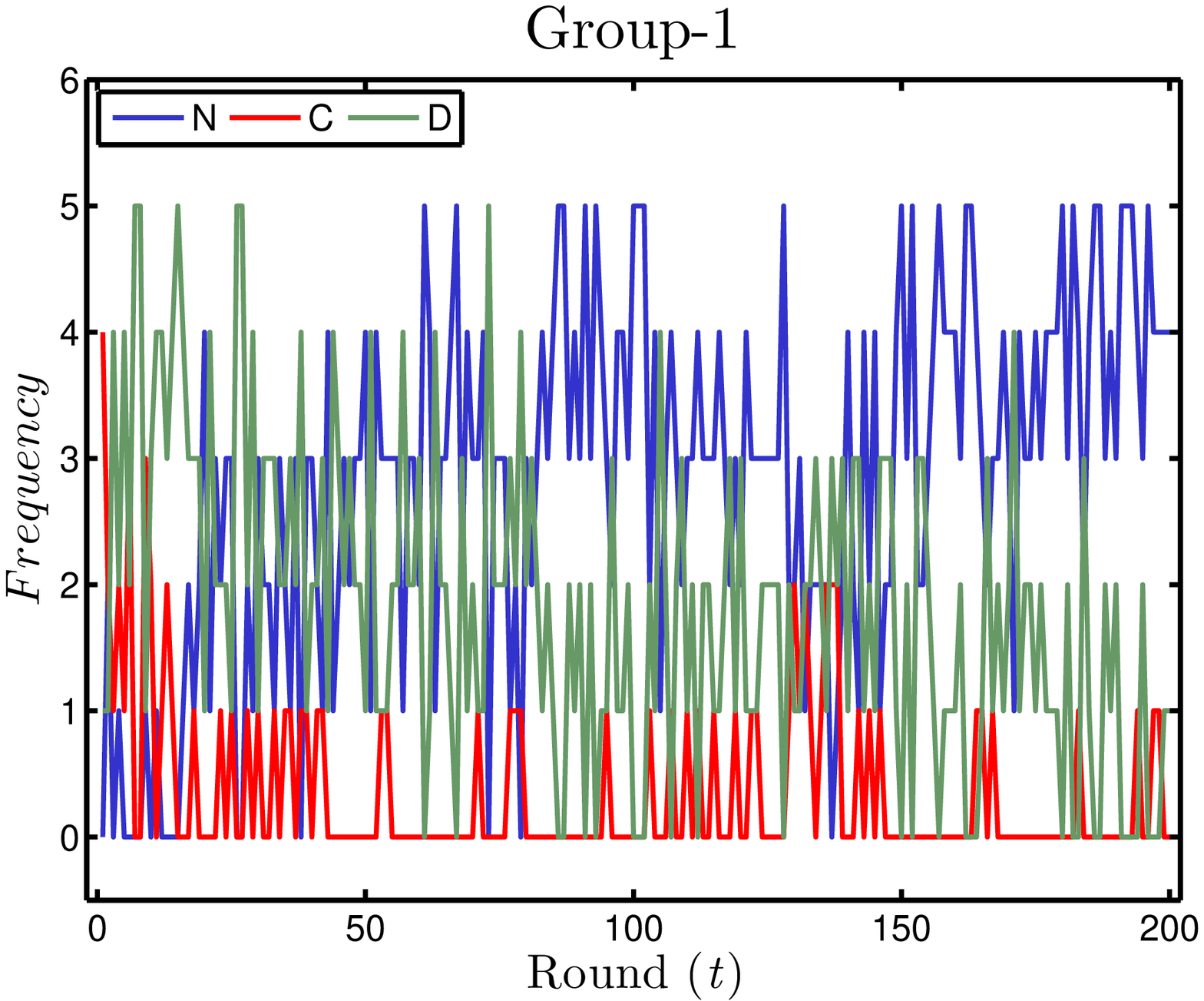}
\includegraphics[angle=0,width=2cm]{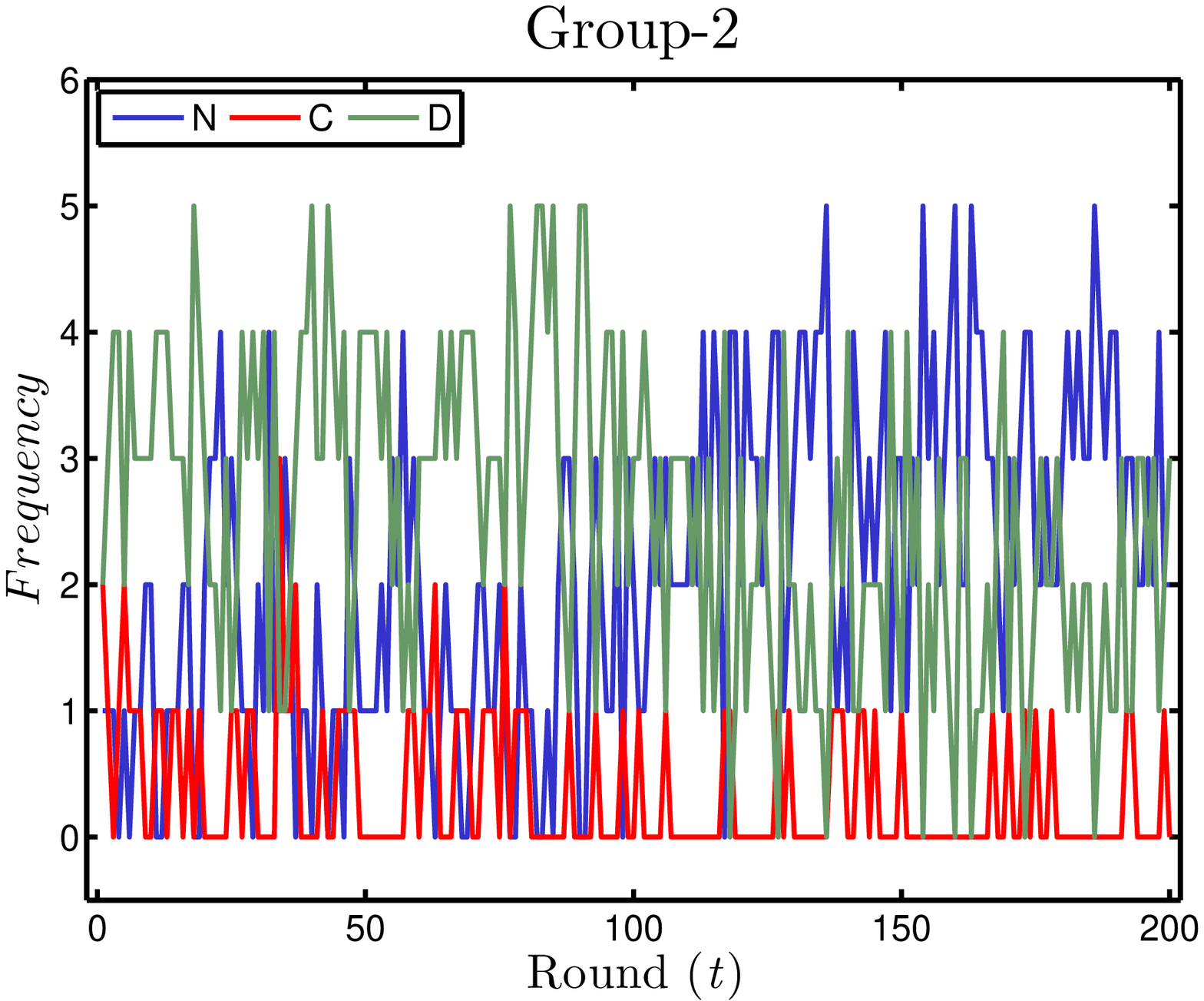}
\includegraphics[angle=0,width=2cm]{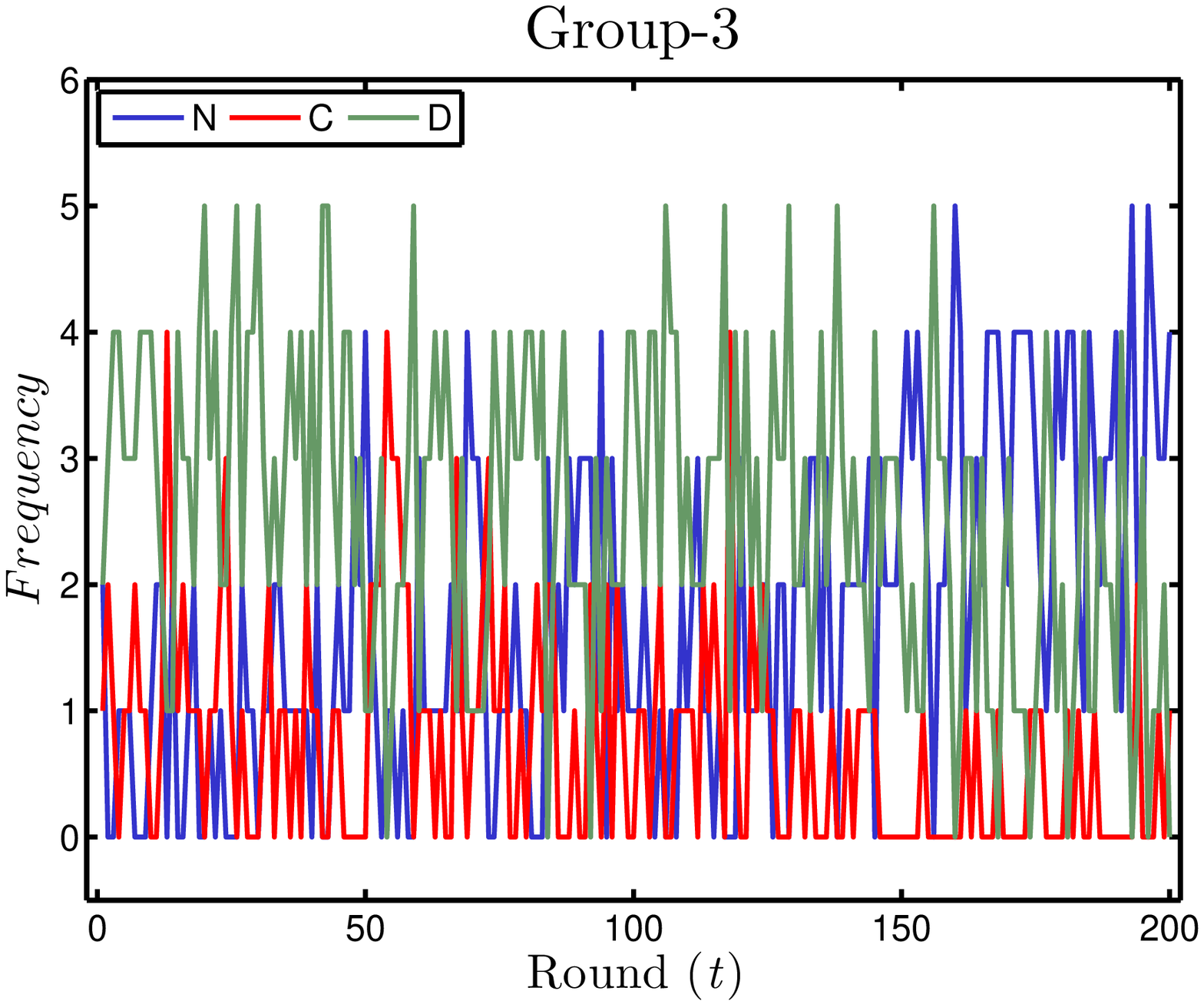}
\includegraphics[angle=0,width=2cm]{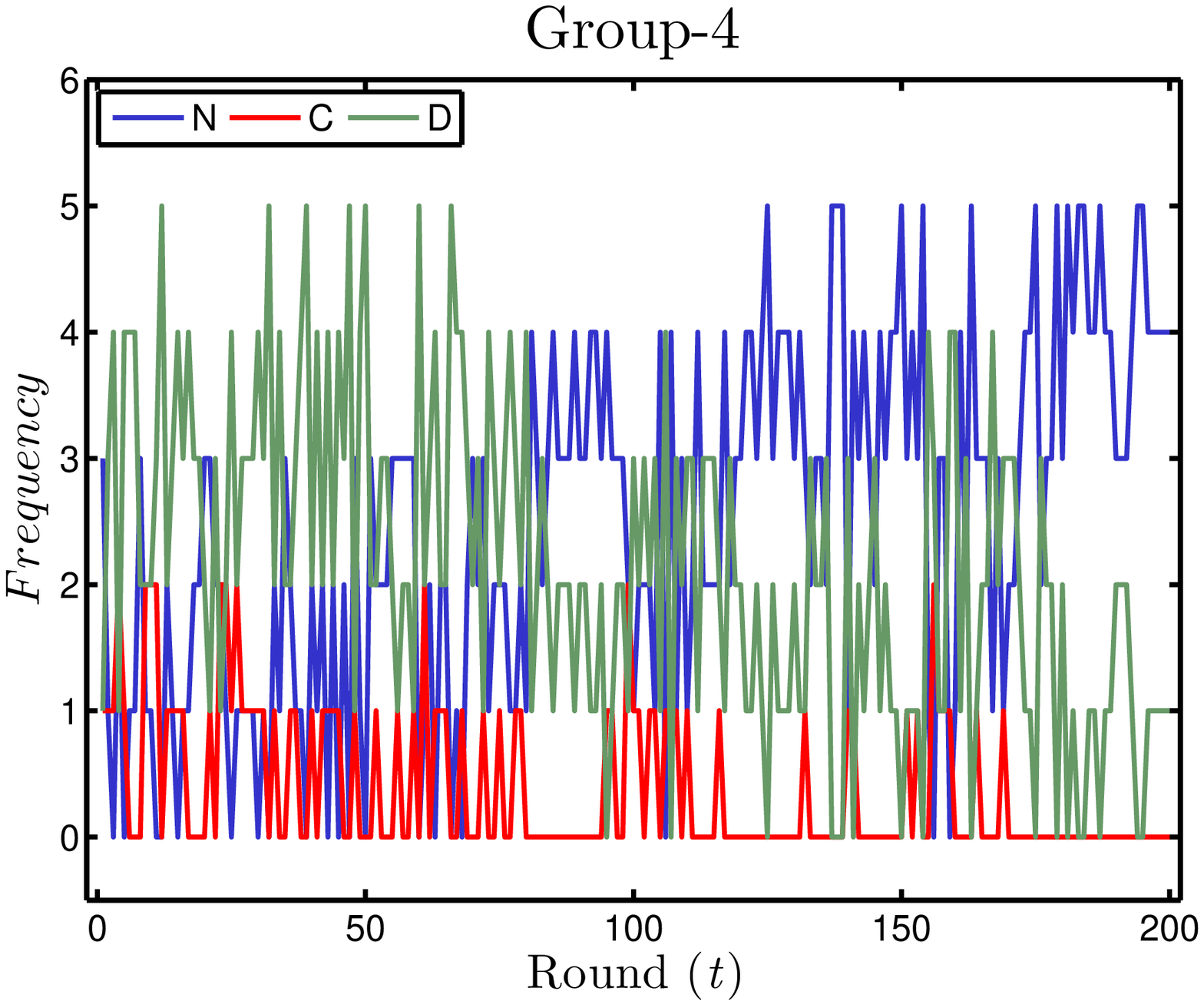}
\includegraphics[angle=0,width=2cm]{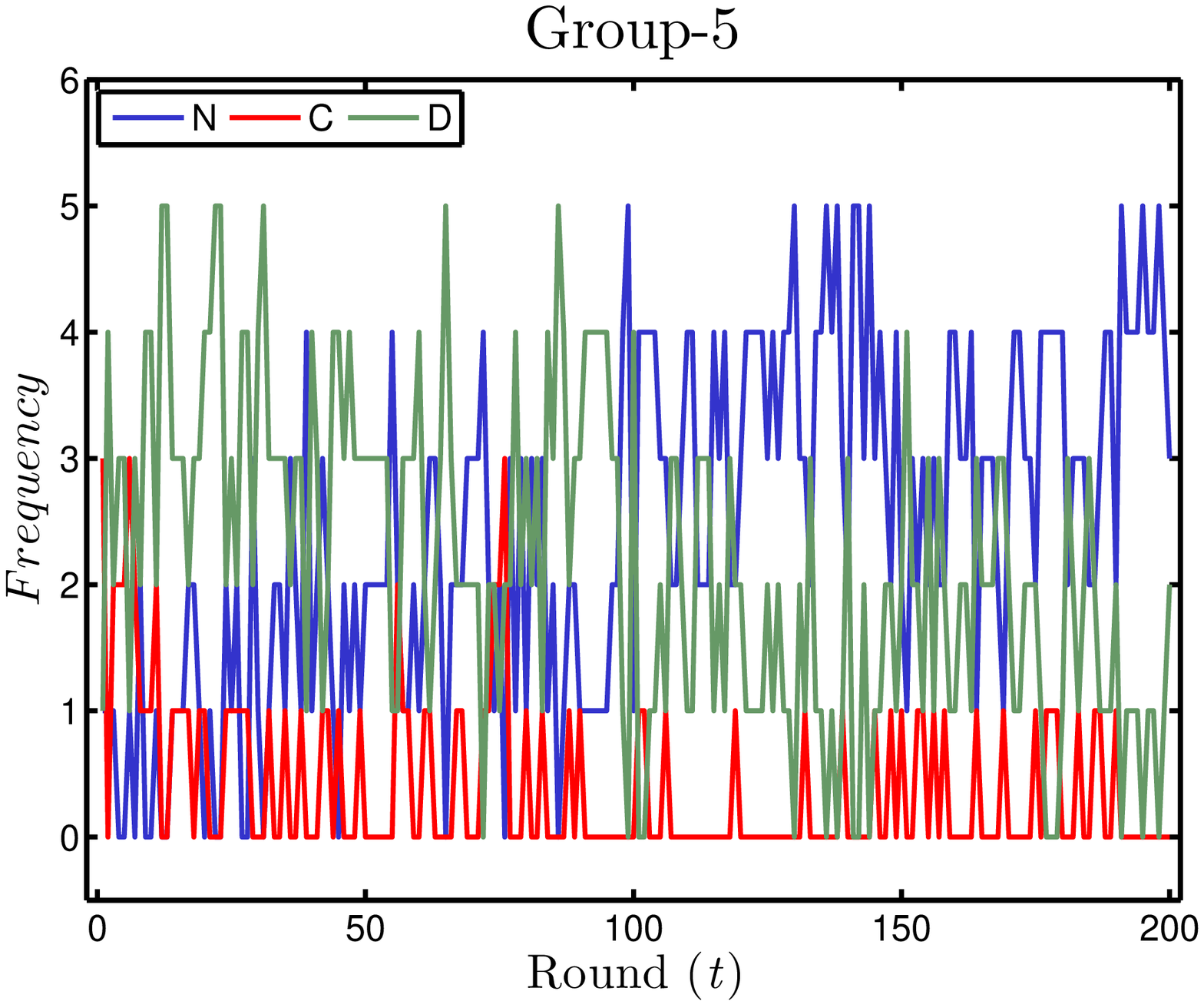}
\includegraphics[angle=0,width=2cm]{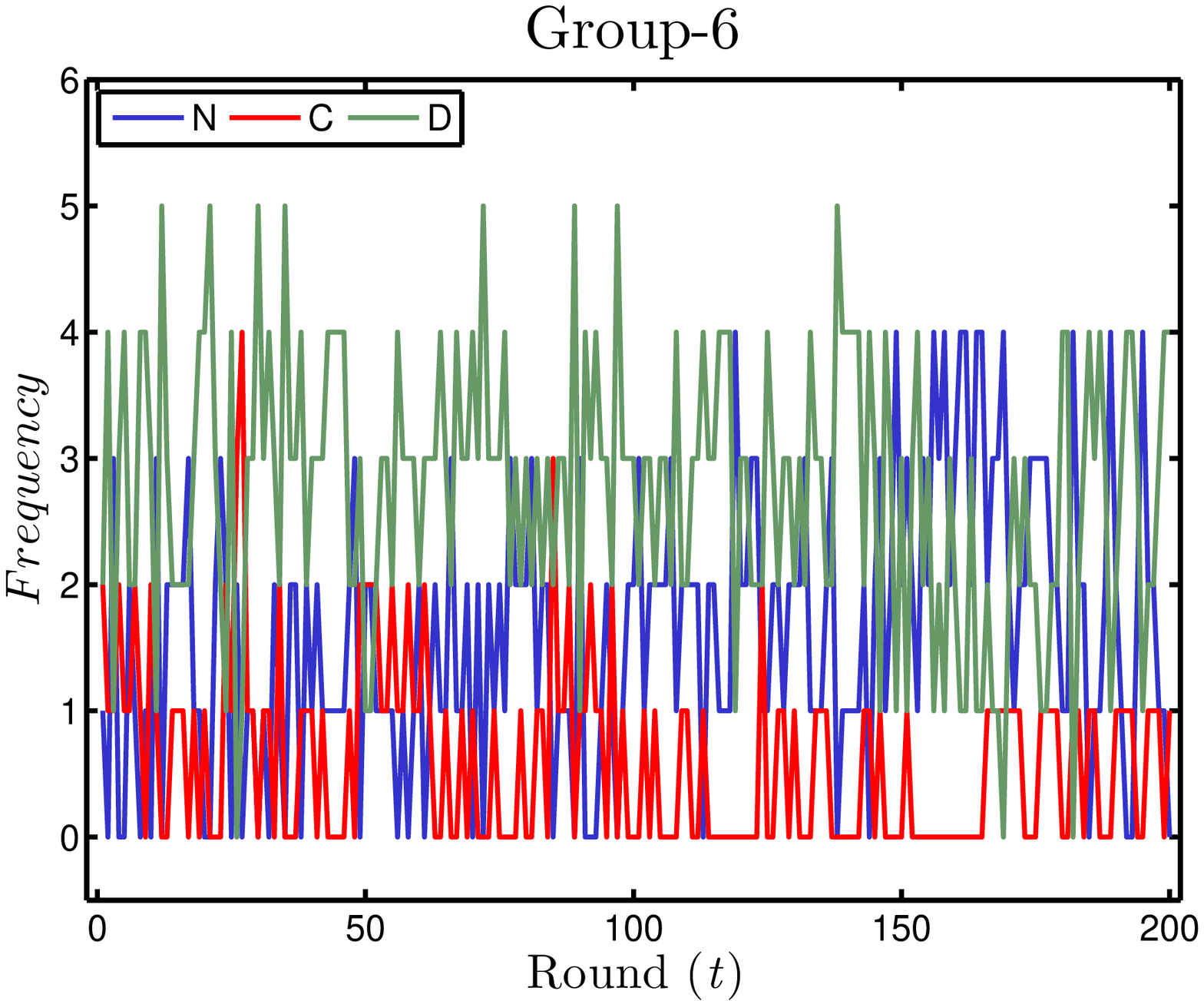}
\caption{Each group's strategy use over time. The graphs in the first and second rows are of T1 and the graphs in the third row are of T2.\
\label{fig:Each group's strategy use over time}}
\end{figure}

However, we should notice that the above graph is the synthetic (average) trend of strategy use over the time, which implies that the curve may cover the details of the strategy change of each group, only if the different groups have same frequency of oscillation and happen to move synchronously. We, while analysing the dynamic pattern, we must use the group level data independently. Every group's strategies use over the time is illustrated in Figure~\ref{fig:Each group's strategy use over time}.

The trend of strategy use of group is roughly similar to that of the overall trend, but the fluctuation of each strategy use is more obvious. Especially, the cooperation strategy (red line) emerges occasionally after a long declining period and all strategies fluctuate abruptly till end. This inspires us to detect the regular pattern (the law) round by round.

\subsection{The mean velocity of state change}

Evolutionary game theory assumes that payoff of a strategy determines the growth rate of its frequency within the population~\cite{sandholm2010population}. The populations can evolve, in the sense that the frequencies $x_{i}$ change with time; and let the state $x(t)$ depend on time, and denote by $\dot{x}_{i}(t)$ the velocity with which $x_{i}$ changes, i.e. $\dot{x}_{i}=dx_{i}/dt$ (~\cite{sigmund2010calculus} p31). Specifically, for example, the replicator dynamics~\cite{taylor1978evolutionary} gives the velocity of the frequencies $x_{i}$ of the strategies $i$ by

\begin{equation}
\dot{x}_{i}=\sum_{j} x_{i}x_{j}(P_{i}-P_{j})=x_{i}(P_{i}-\bar{P})
\label{eq:replicator}
\end{equation}
where $\bar{P}=\sum x_{j}P_{j}$ is the average payoff in the population.

\begin{figure}
\centering
\includegraphics[angle=0,width=4cm]{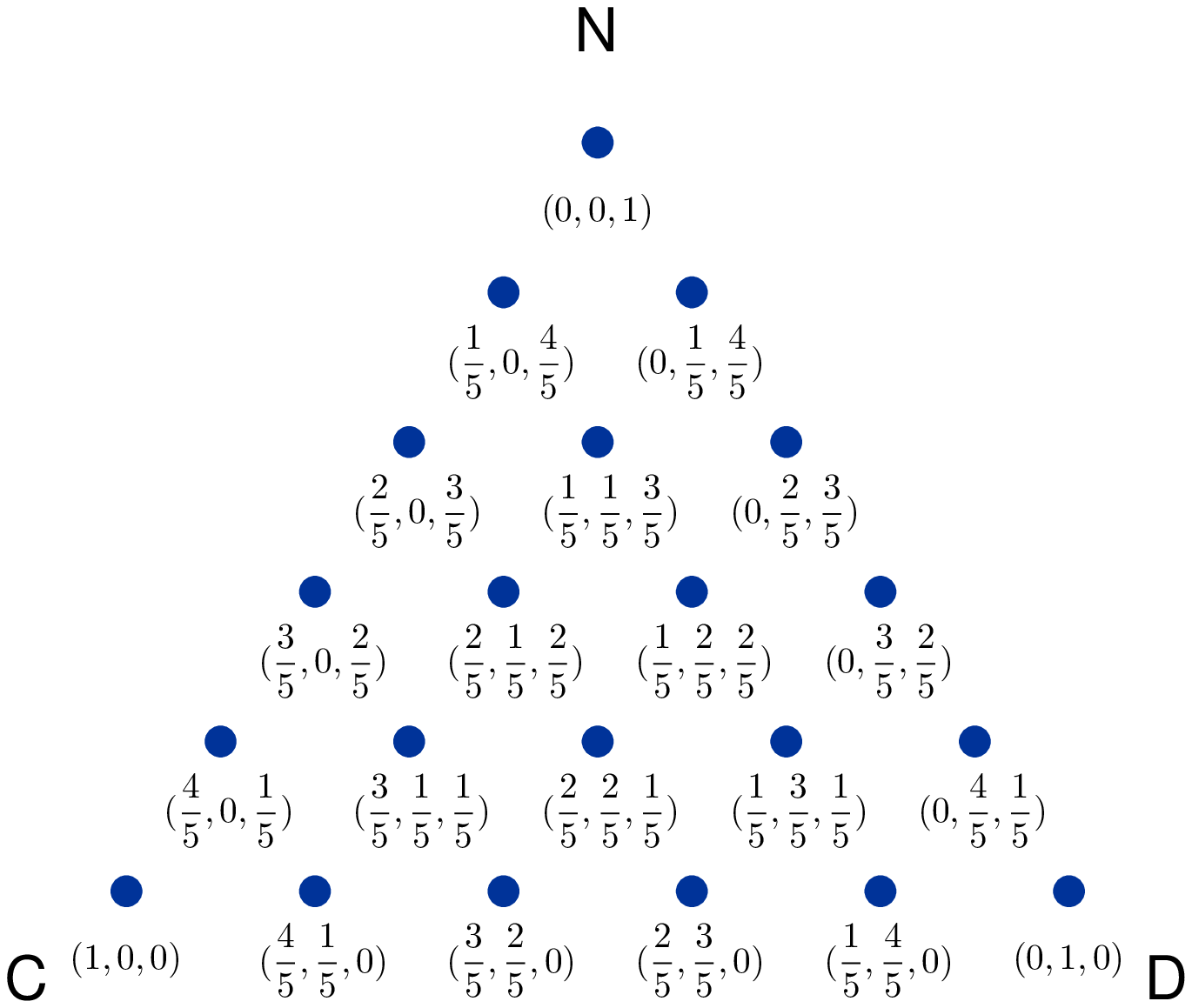}
\includegraphics[angle=0,width=4cm]{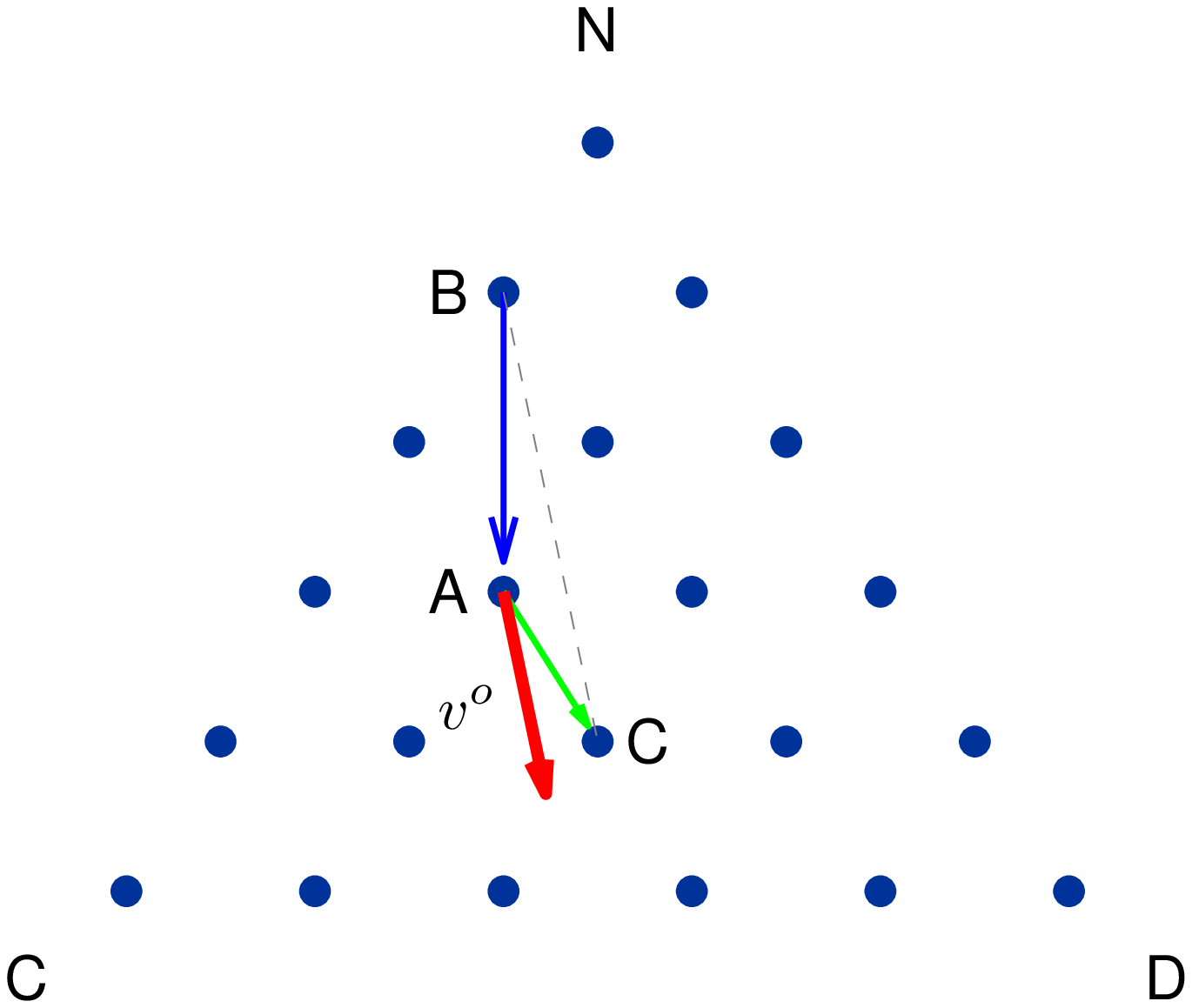}
\includegraphics[angle=0,width=4cm]{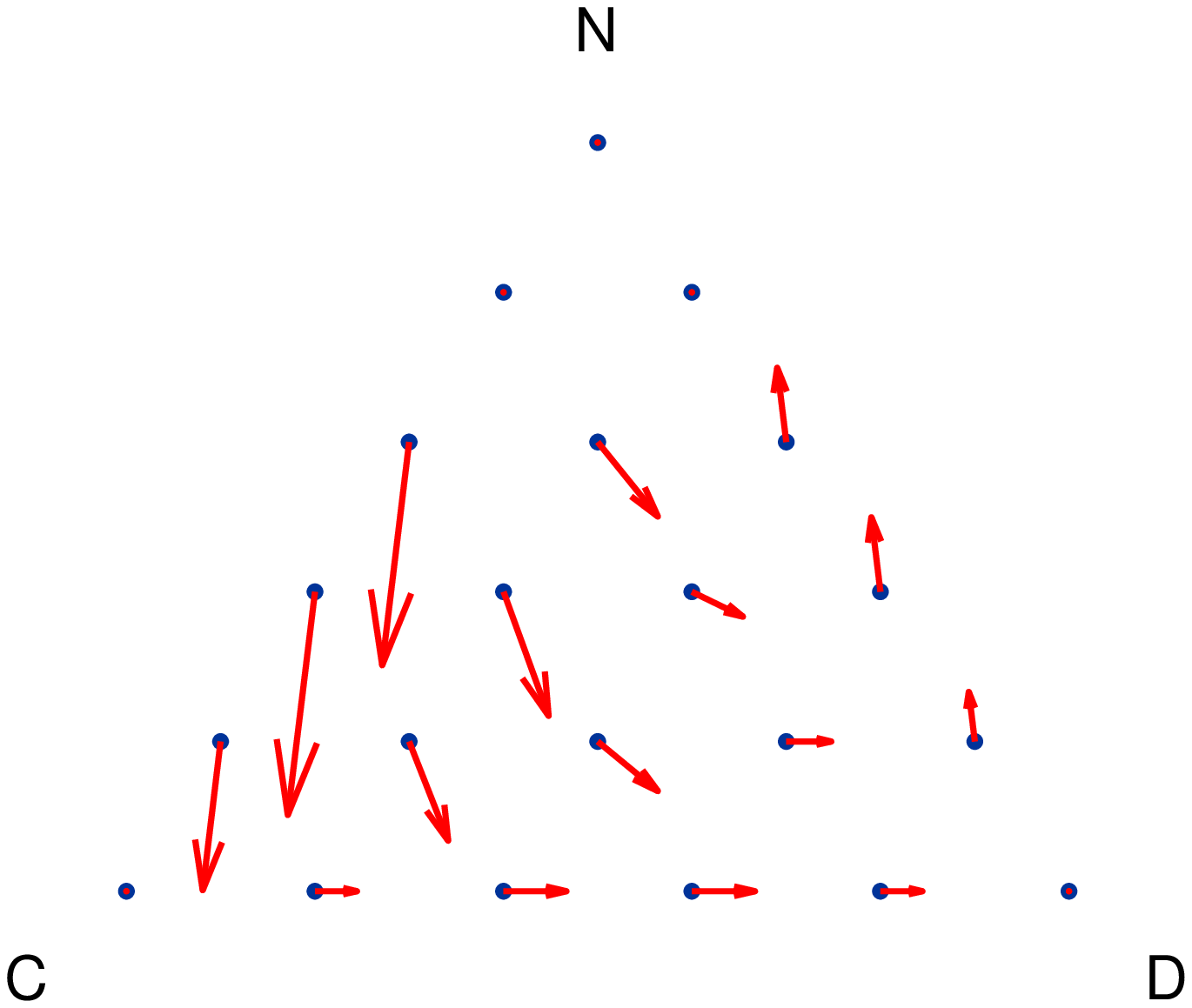}
\caption{State space and velocity schematic diagram.\
\label{fig:State space and velocity schematic diagram} The left panel is state space for the public goods game in which each point presents a social state. The three numerals under every state represent the frequencies of strategy C, D and N, respectively. The middle panel is a schematic diagram about velocity. At time $t$, the group is at state A, i.e. $x^{o}=(\frac{2}{5},\frac{1}{5},\frac{2}{5})$, at time $t-1$, the group is at state B, i.e., $x^{-}=(\frac{1}{5},0,\frac{4}{5})$, and at time $t+1$, the group is at state C, i.e. $x^{+}=(\frac{2}{5},\frac{2}{5},\frac{1}{5})$. The right panel is an example of evolutionary velocity of strategies given by replicator dynamics.}
\end{figure}

In every period, the population must be in a certain state which is described by a combination of strategies' frequencies~\cite{Sandholm2011}. Here, in CDN game, the social state is denoted by $x_{ncd}$ or $( x_{C},x_{D},x_{N})$, where, $x_{C}$, $x_{D}$, and $x_{N}$ represent the proportion of players choosing strategy C, D and N, respectively. For example, in a society of five, the social state is $(\frac{1}{5},\frac{2}{5},\frac{2}{5})$, if one person chooses C, two persons choose D and the remaining two choose N. Fig.~\ref{fig:State space and velocity schematic diagram} (left) represents the social states in the state space of the CDN game.

In physics, velocity is the measurement of the rate and direction of change in position of an object. And the instantaneous velocity is always tangential to trajectory. The velocity of a society evolution is a vector of all strategies' change. Theoretically, an evolutionary dynamics equation has provided the velocity in the full state space. Following the velocity notion above, Xu and Wang~\cite{xu2010bertrand,xu2011evolutionary,xu2011observation} introduced an empirical method to measure the velocity of strategy density change in the state space. Given a certain state $x^{o}=(x_{c},x_{d},x_{n})$ in the state space as an object of observation, suppose that the group is just right at the state $x^{o}$ at time $t$. For this observation $x^{o}$, there is a state $x^{+}$ in the next round $t+1$ and a state $x^{-}$ in the previous round $t-1$. The empirical velocity at a state $x^{o}$ is given by:

\begin{eqnarray}
v^o=[(x^{+} - x^{o})+(x^{o}- x^{-})]/(2\bigtriangleup t) = (x^{+}- x^{-})/(2\bigtriangleup t),
    \label{EqVone1}
\end{eqnarray}
where $\bigtriangleup t=1$ indicates the time duration between two rounds in experiment, i.e. the time of one round. The key point of the measurement here is that during the calculation of the velocity, the backward change connot be ignored. The velocity is the composition of forward velocity $v^{+}=(x^{+} - x^{o})/\bigtriangleup t$ and backward velocity $v^{-}=(x^{o}- x^{-})/\bigtriangleup t$, $v^o=(v^{+}+v^{-})/2$.

From time to time, the given state $x^{o}$ may be passed many times. The average velocity $\bar{v}^{o}$ at this state $x^{o}$ is given by:

\begin{eqnarray}
\bar{v}^{o}=\sum_{o} v^o/\Omega,
    \label{EqVone2}
\end{eqnarray}
where $\Omega$ is the frequency of observed $v^o$.

\begin{figure}
\centering
\includegraphics[angle=0,width=6cm]{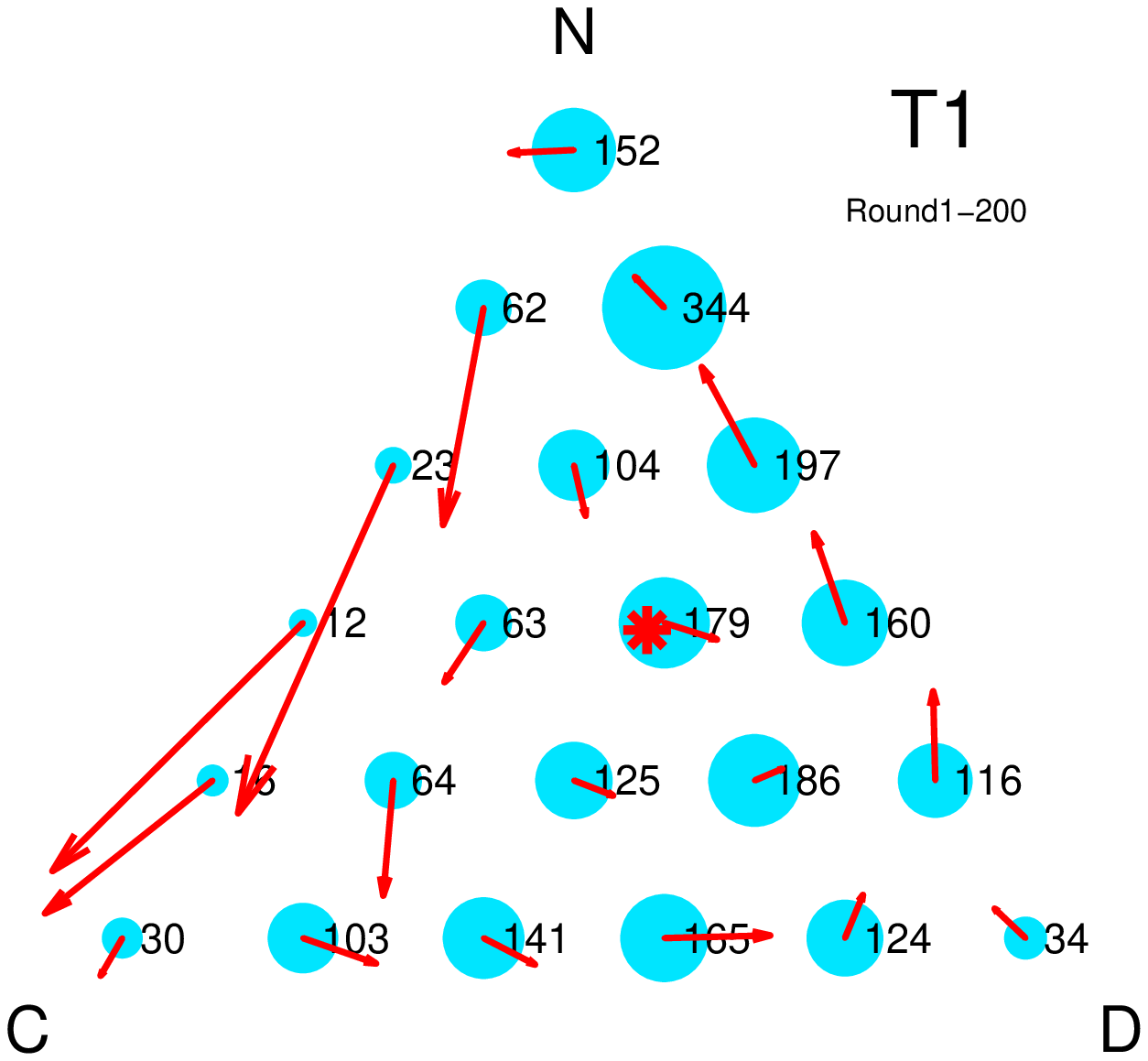}
\includegraphics[angle=0,width=6cm]{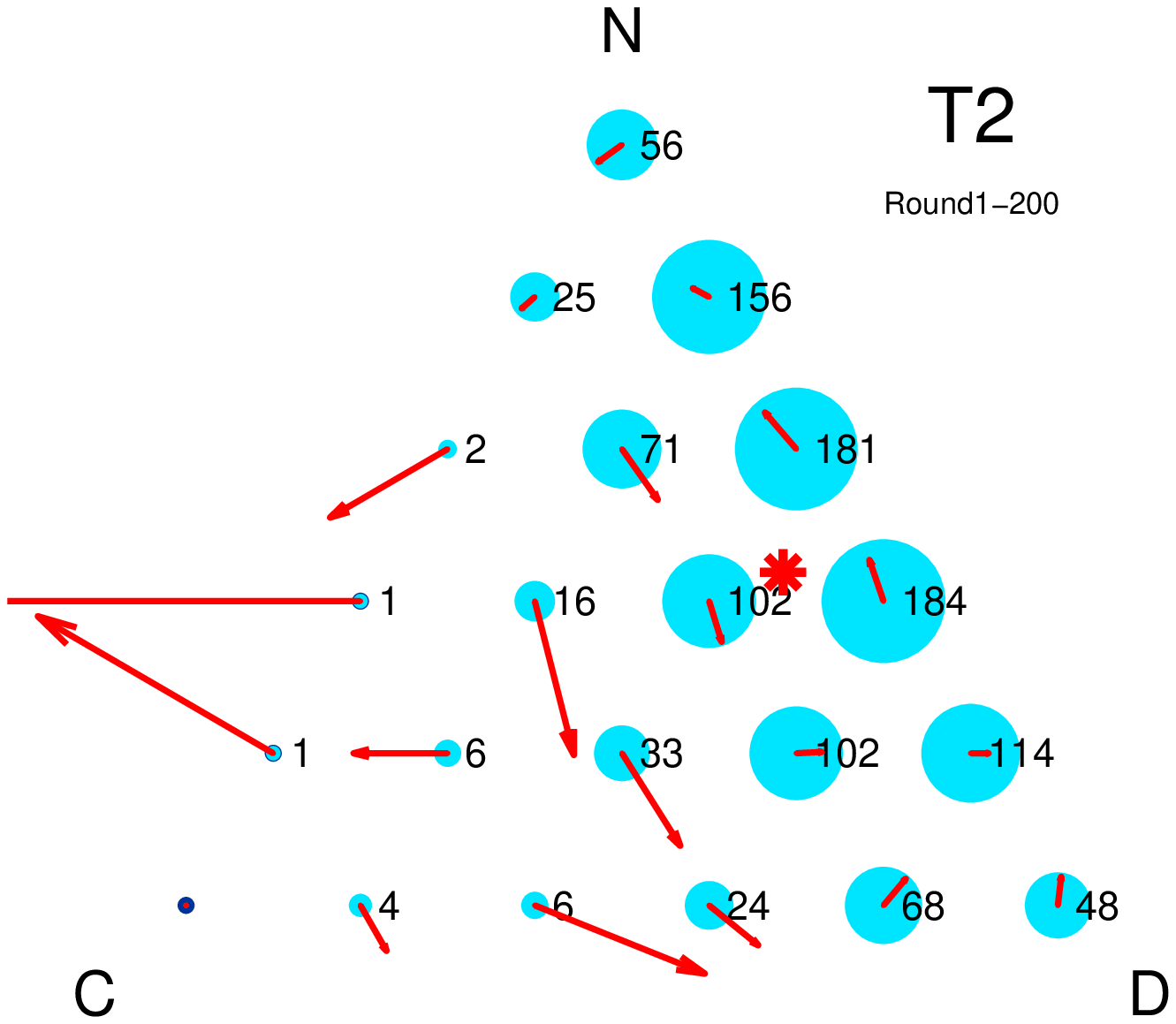}
\caption{The mean velocity of state change. The left panel is Treatment 1 and right one is Treatment 2. The red arrow at every state is the mean velocity of the state change which is calculated by Eq.\ref{EqVone2}. While the number in the blue circle at every state is the observed frequency of that state, and the size of the circle is relative to the number. The red star is the mean observed state of all groups of all rounds.
\label{fig:velocity of strategy density change}}
\end{figure}

Now it is easy to get the velocity for every state, because all the conditions are getting the state data. According to the individual's choice, we calculate the state vector for every group of every round, then according the Eq.~\ref{EqVone2}, we get the velocity at every state (Figure.~\ref{fig:velocity of strategy density change}).

Obviously, the cyclic dynamic patterns are clear in both treatments, especially in treatment 1. The directions are consistent with the evolutionary dynamics, for example the replicator dynamics (shown in Figure~\ref{fig:State space and velocity schematic diagram}). Magnitude and direction of velocity to compare empirical velocity with the theoretical velocity at every state are used. Each velocity has its component $v_{x}$ and component $v_{y}$. The magnitude $\|\bar{v}^{o}\|$ of velocity as well as direction, which can be described by the intersection angle $\alpha$ from $v_{x}$ to $\bar{v}^{o}$ are given by equation:

\begin{eqnarray}
 \|\bar{v}^{o}\|=\sqrt{\bar{v}^2_{x}+ \bar{v}^2_{y}}\\
\alpha=arctan\frac{\bar{v}_{y}}{\bar{v}_{x}}
\label{eq:alpha}
\end{eqnarray}

The replicator dynamics equation is used as an example to get the theoretical velocity and its corresponding magnitude $\|v\|$ and angle $\alpha$. For each state, we get the empirical velocity, that means, total 21 outcomes. However, the replicator dynamics only provides 16 outcomes, we get no information on the state $(1,0,0)$, $(\frac{1}{5},0,\frac{4}{5})$, $(0,0,1)$, $(0,\frac{1}{5},\frac{4}{5})$, and $(0,1,0)$. So, liner regression is adopted for these 16 observations. The coefficients of magnitude $\|\bar{v}^{o}\|$ are 0.135 ($p$=0.000, $Adj$-$R^2$=0.821) in T1 and 0.130 ($p$=0.001, $Adj$-$R^2$=0.554) in T2, respectively, while the coefficients of angle $\alpha$ are 0.497 ($p$=0.037, $Adj$-$R^2$=0.224) in T1 and 0.290 ($p$=0.080, $Adj$-$R^2$=0.145) in T2, respectively. Obviously, the results of magnitude are better than those of angle, as well the results of T1 are better than those of T2.

\begin{figure}
\centering
\includegraphics[angle=0,width=3cm]{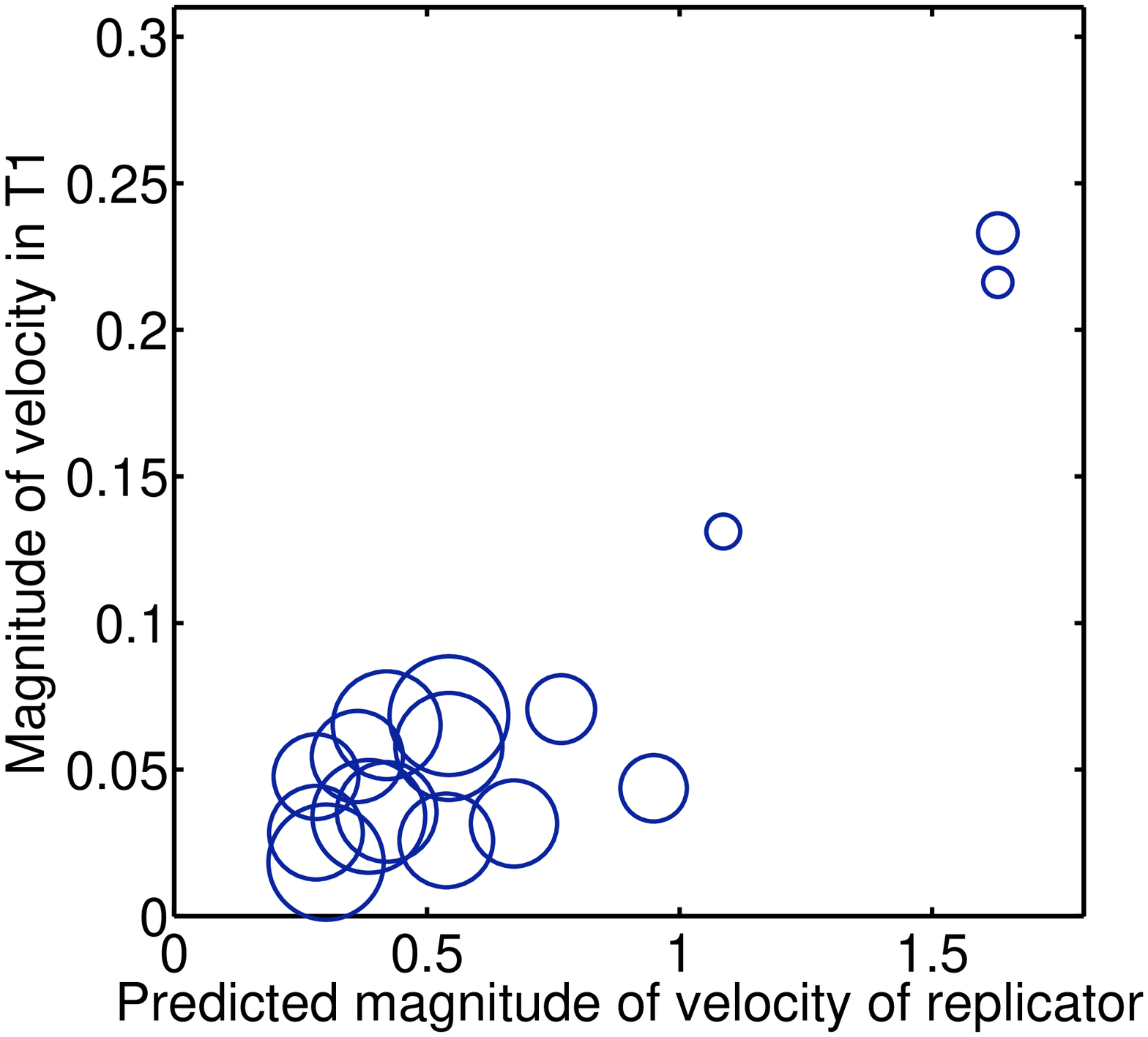}
\includegraphics[angle=0,width=3cm]{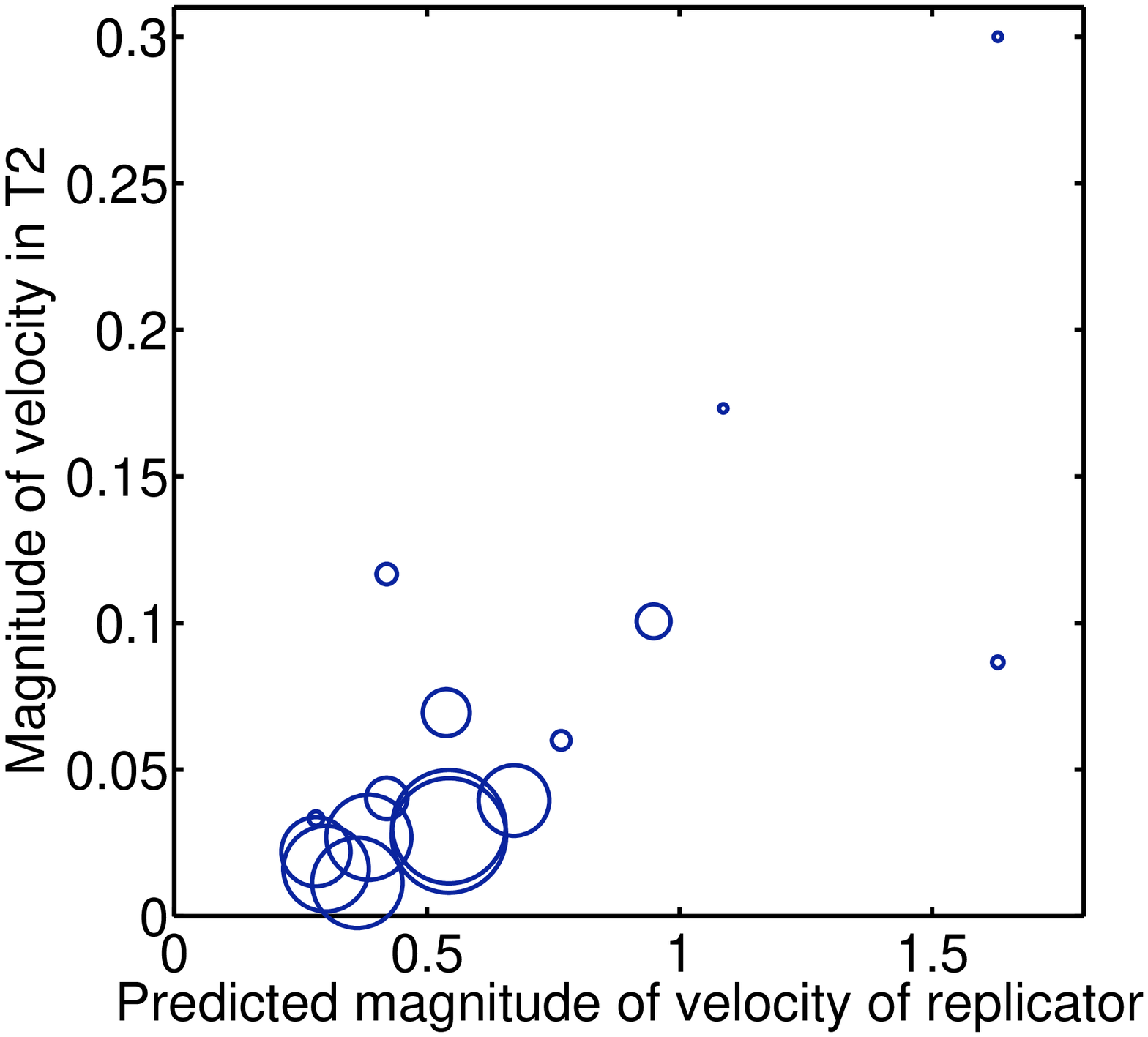}
\includegraphics[angle=0,width=3cm]{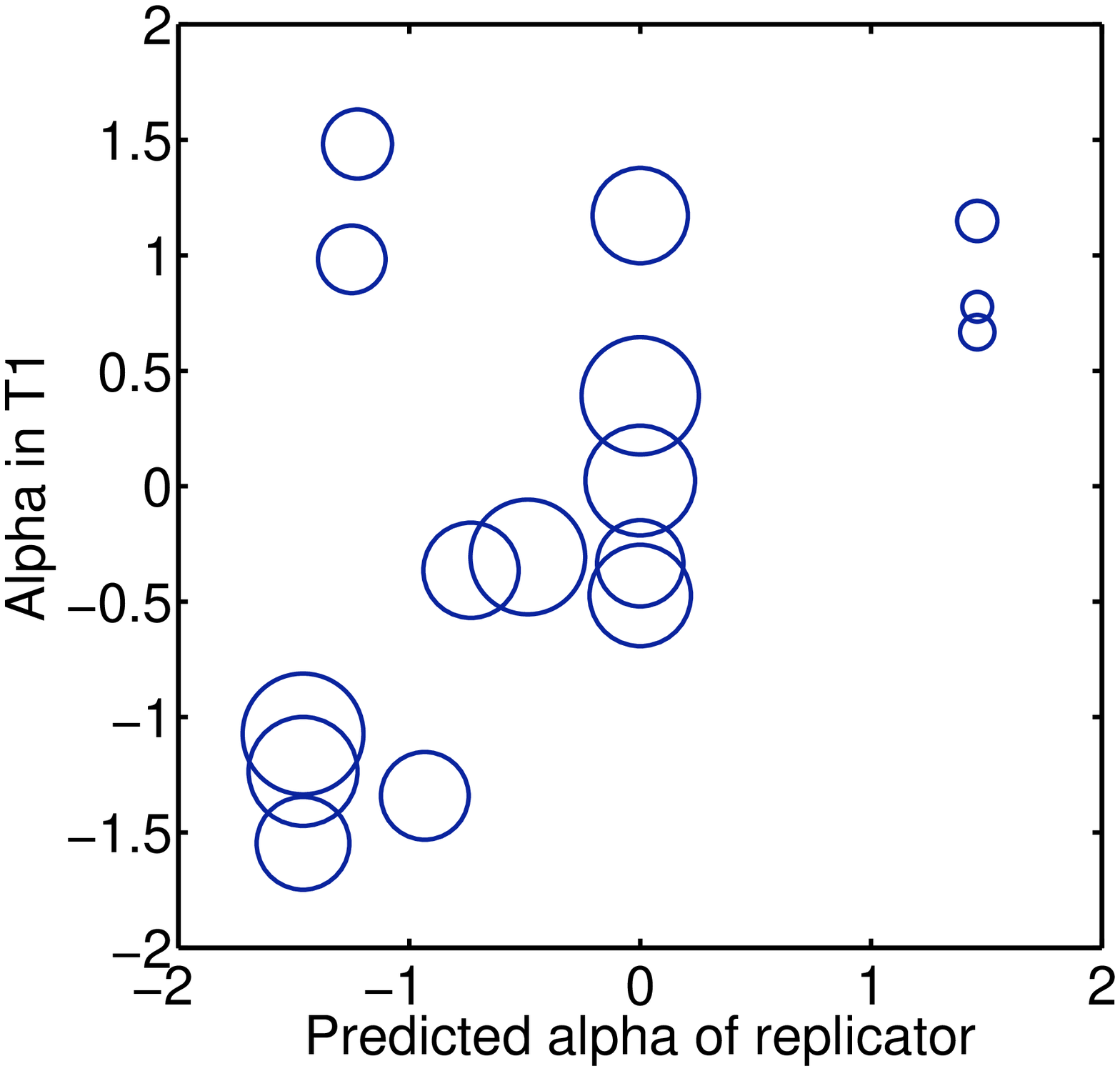}
\includegraphics[angle=0,width=3cm]{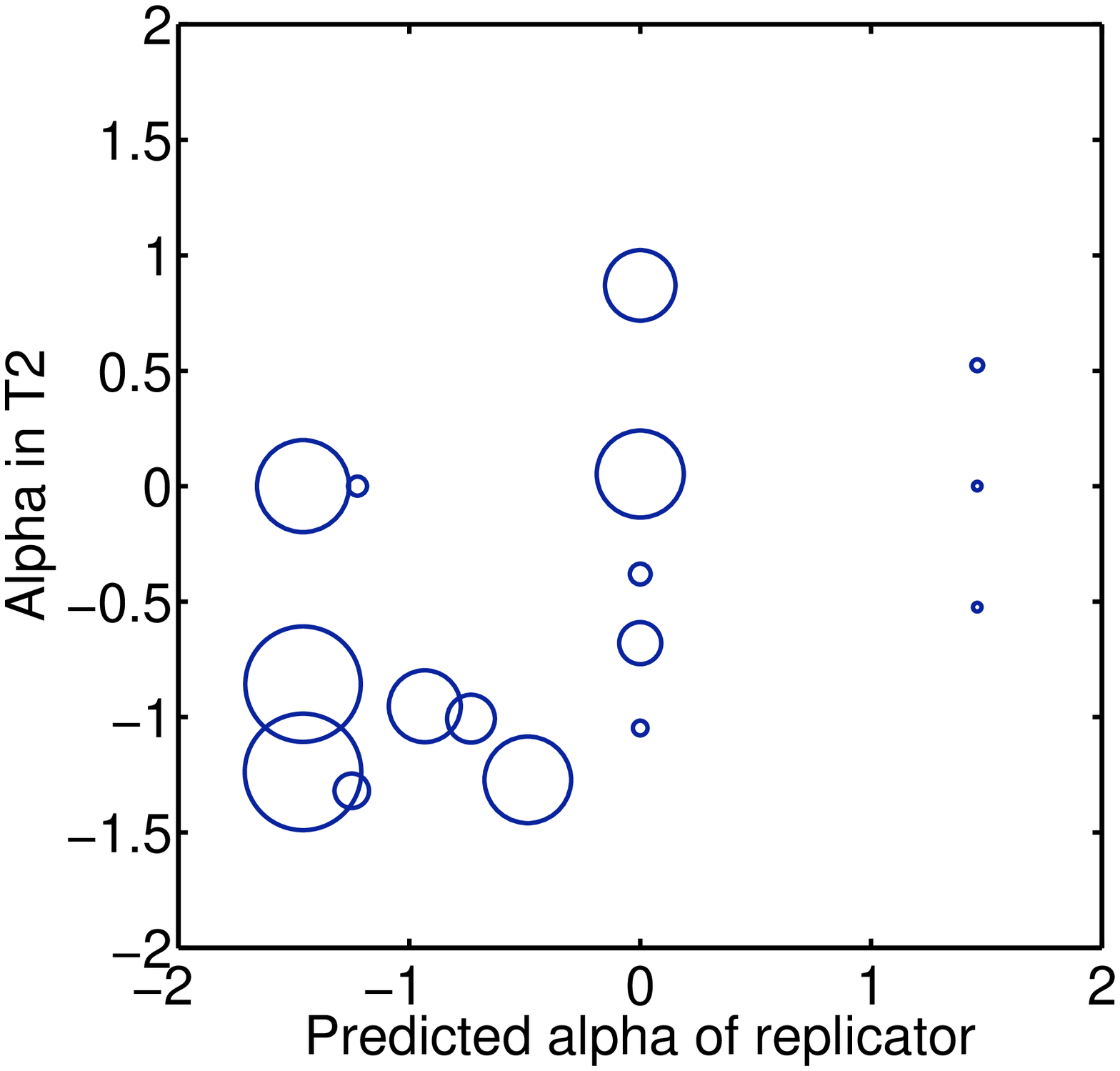}
\caption{Scatter graph. The first two graphs are the results of magnitude and the last two graphs are the results of direction, respectively. The cycle size is related to the frequency of $v^{o}$ at that state.
\label{fig:scatter}}
\end{figure}

As explicated in the Figure~\ref{fig:velocity of strategy density change}, the distribution of state is quite unequal in the space. The distribution of strategies are spread near the DN edge, especially in T2. Hence, the scatter with weights of frequencies at each state is plotted in Figure~\ref{fig:scatter}.

These results seem to implicate that the evolutionary dynamics has ability to capture the real dynamics not only on biosphere, but also on human society.

\subsection{The distribution and velocity over time}

\begin{figure}
\centering
\includegraphics[angle=0,width=3cm]{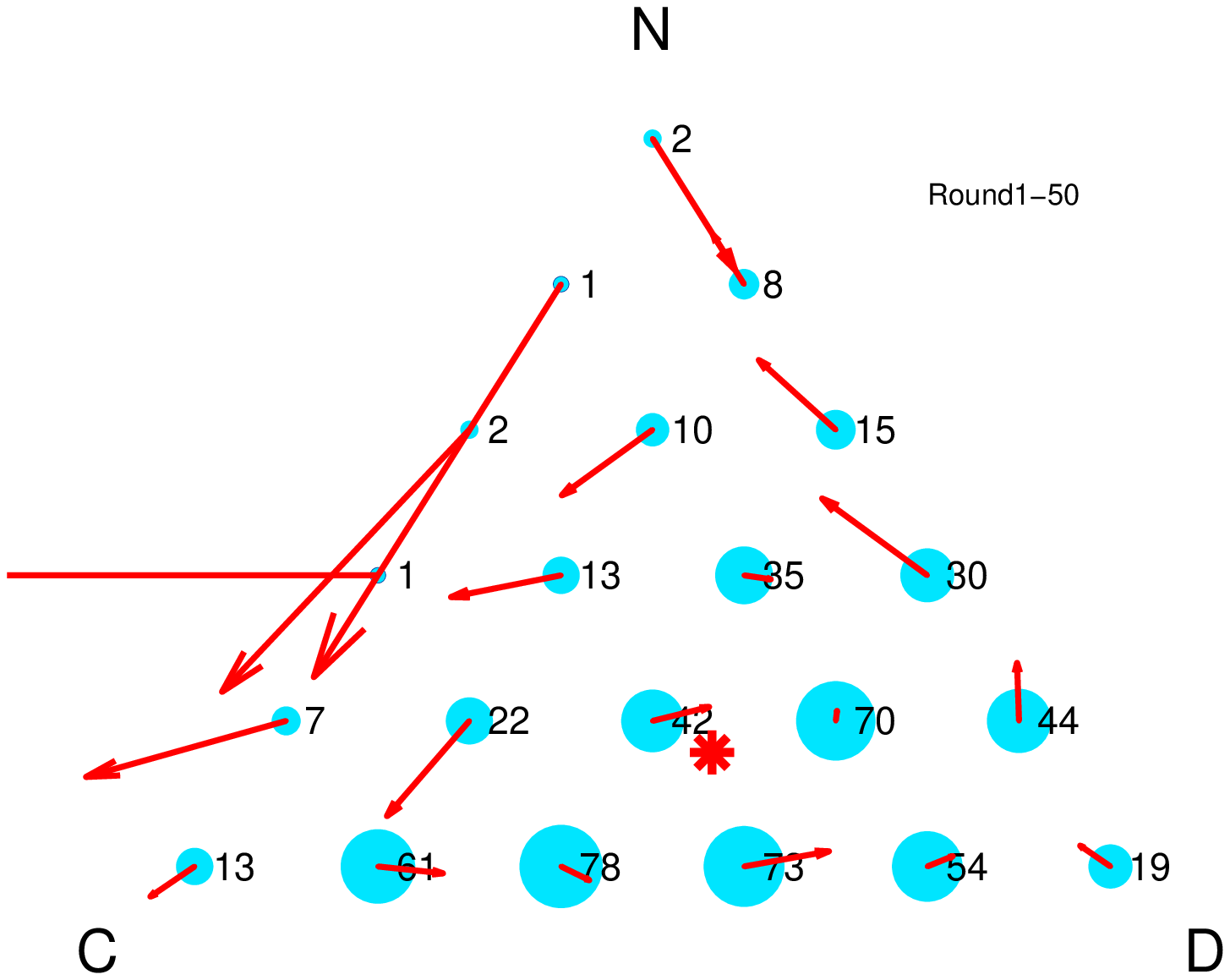}
\includegraphics[angle=0,width=3cm]{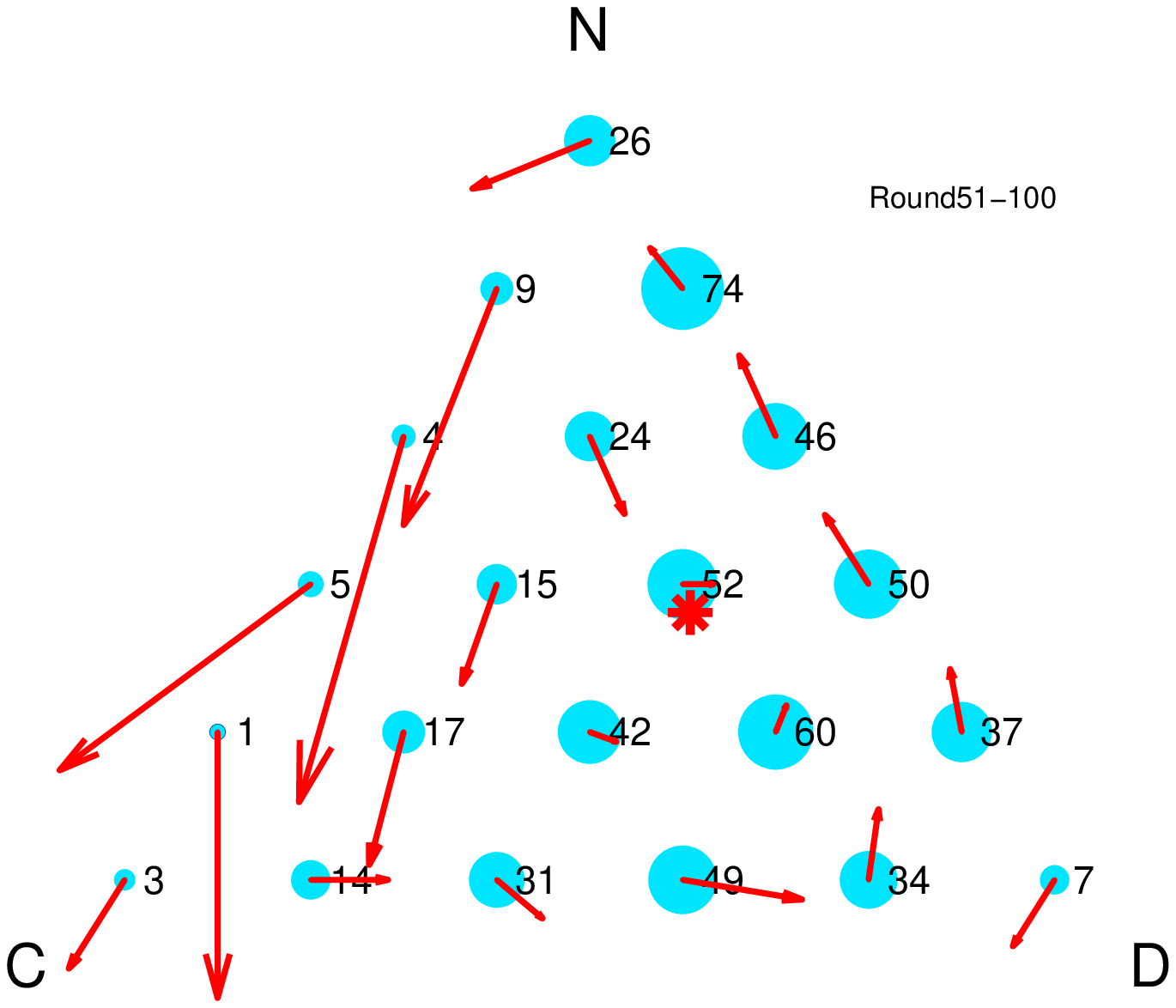}
\includegraphics[angle=0,width=3cm]{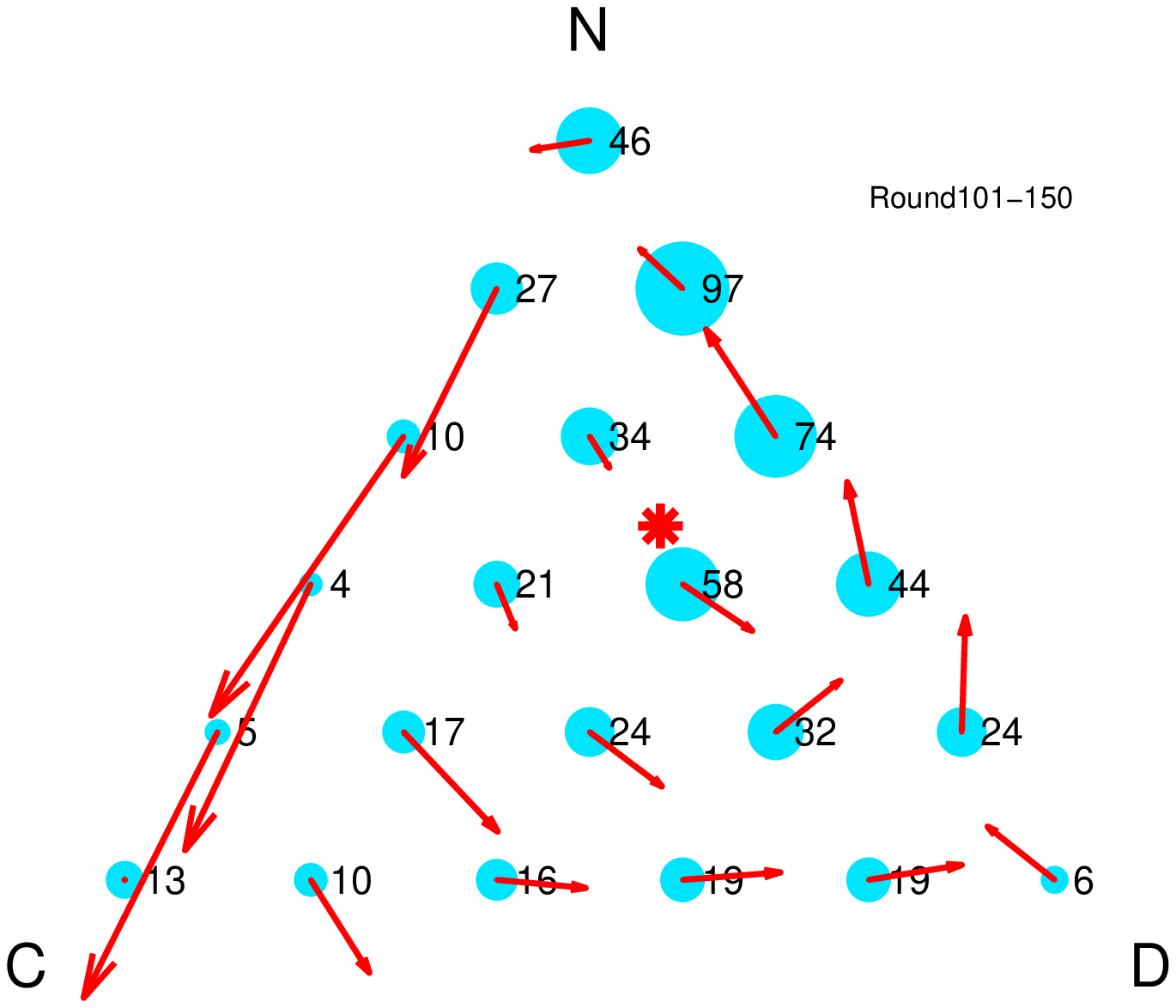}
\includegraphics[angle=0,width=3cm]{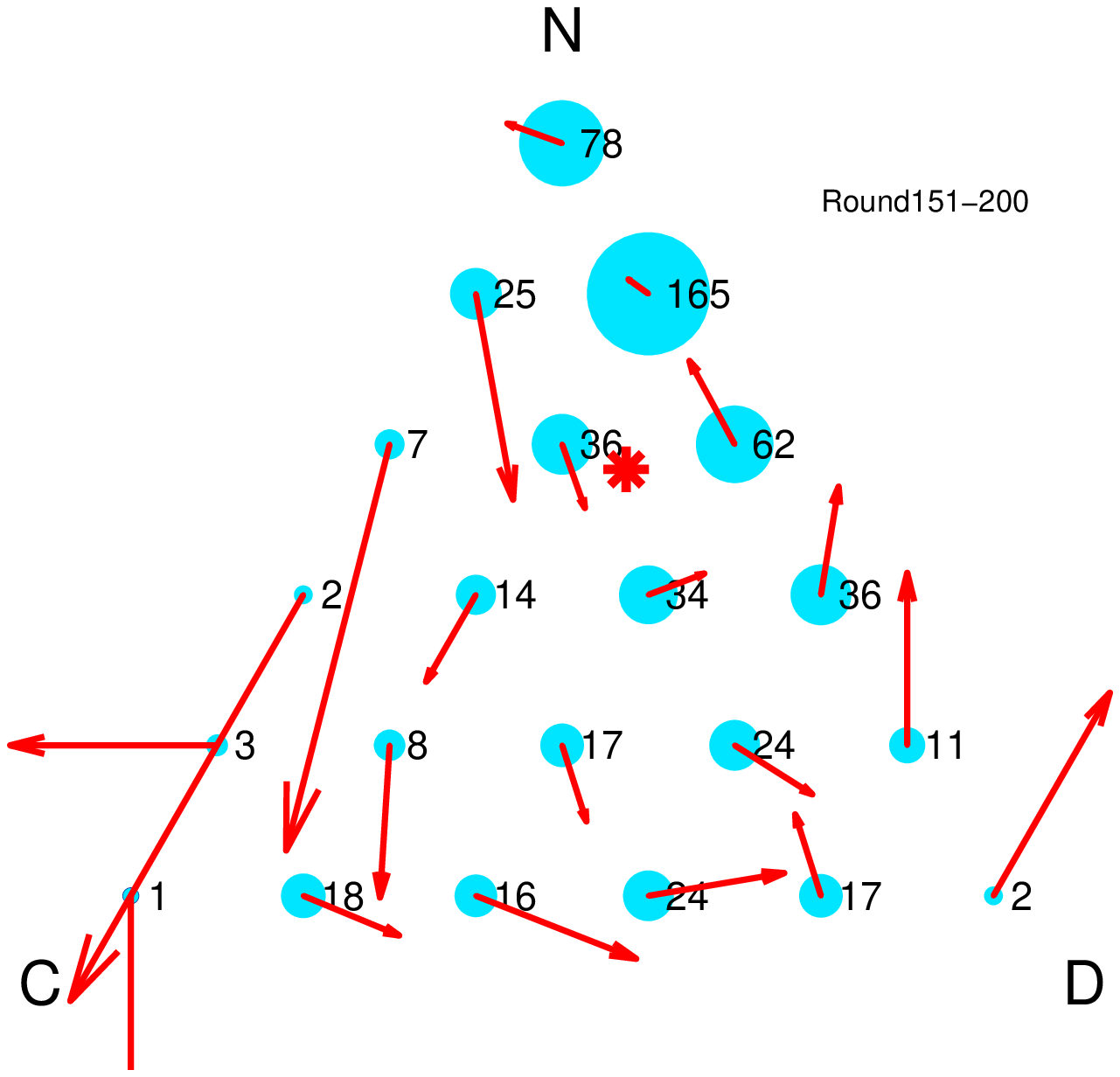}
\includegraphics[angle=0,width=3cm]{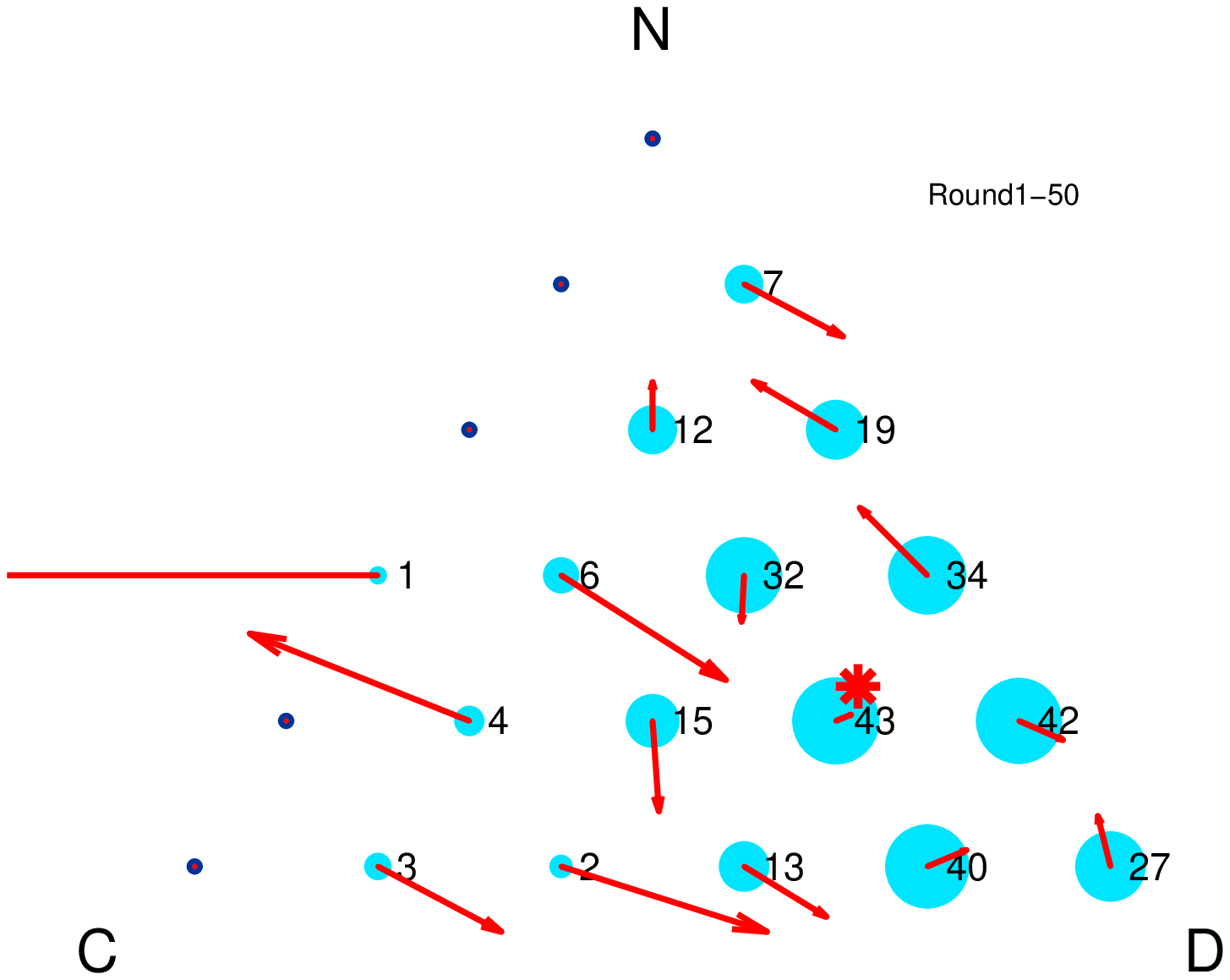}
\includegraphics[angle=0,width=3cm]{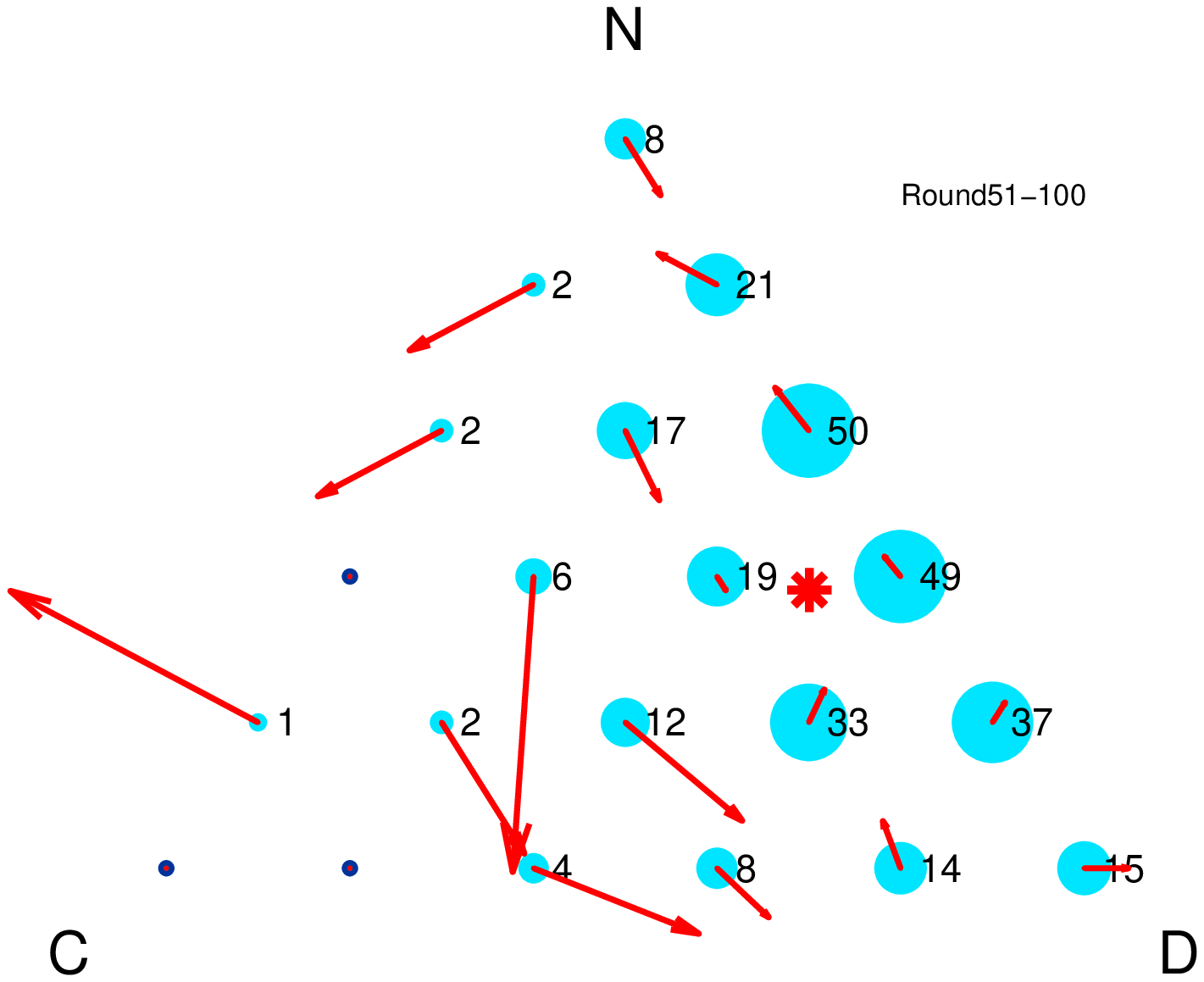}
\includegraphics[angle=0,width=3cm]{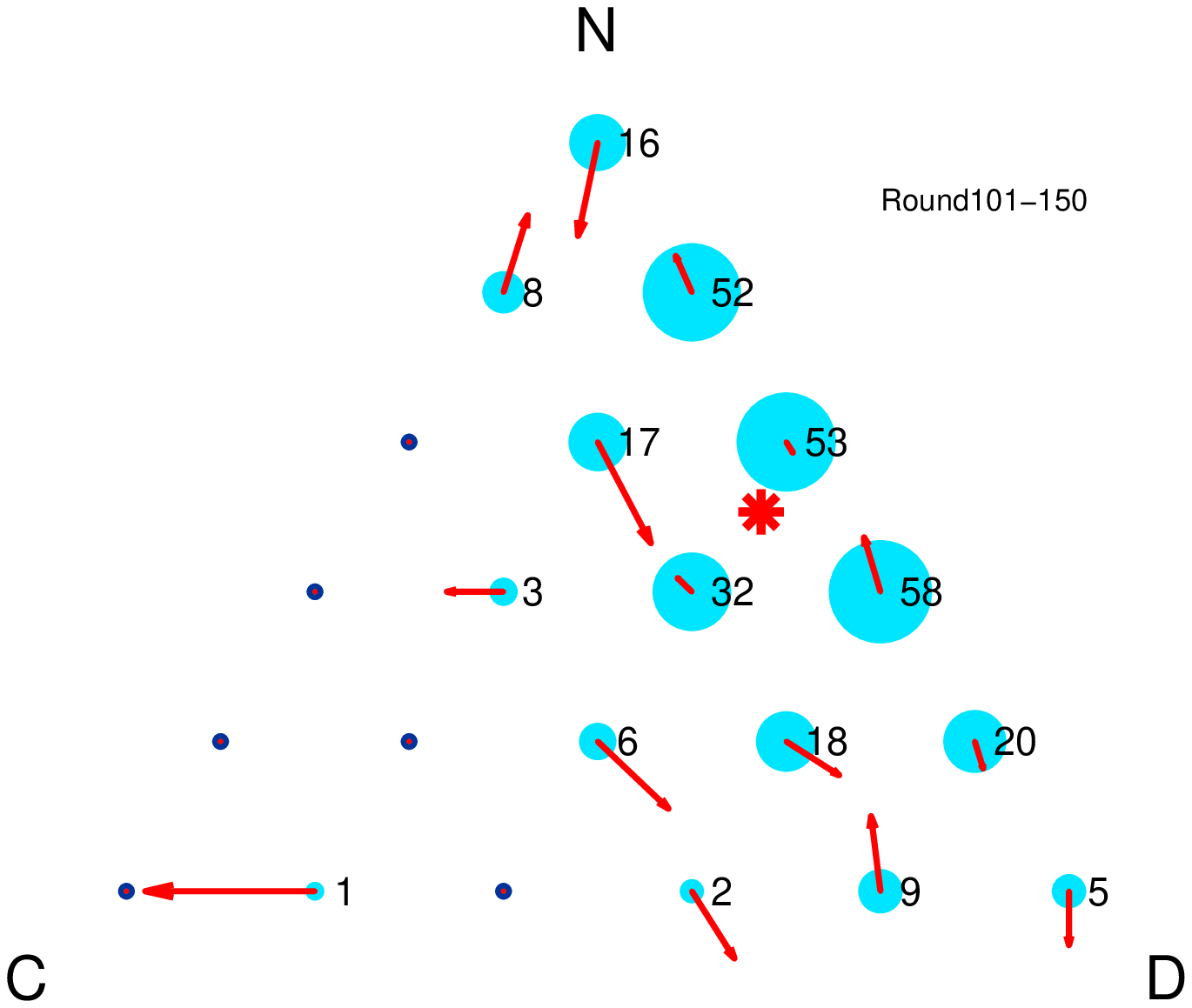}
\includegraphics[angle=0,width=3cm]{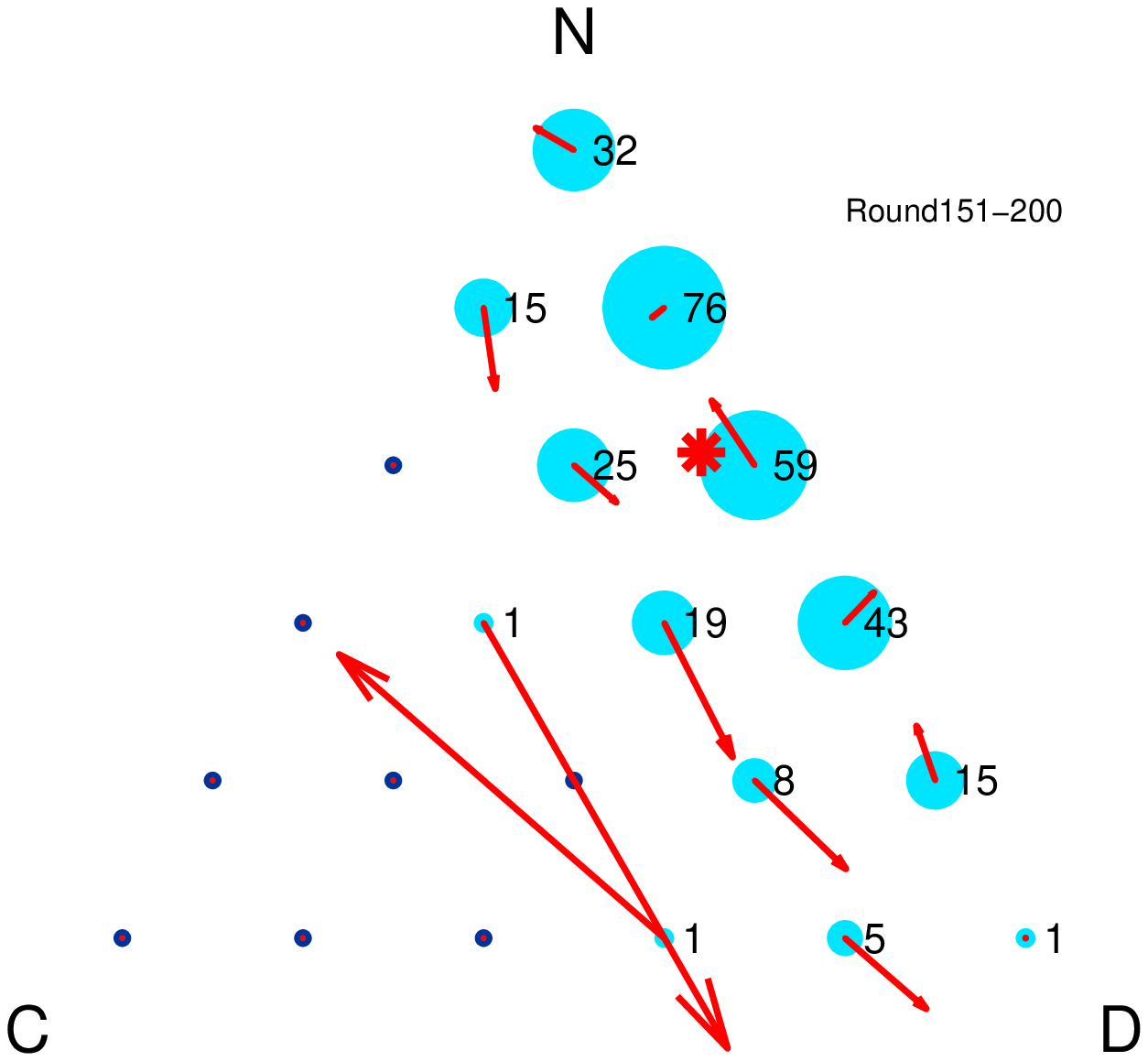}
\caption{The distribution and velocity over time.\
\label{fig:The distribution and velocity over time}}
\end{figure}

The velocity patterns discussed above are the overall results about all 200 rounds, related to the trend of strategies which has been shown in Figure~\ref{fig:strategy ues over time}, where the spiral pattern itself may change over time. To find out whether the cyclic pattern will disappear, when N strategy is used frequently, the 200 rounds are divided successively into four segments, each of them including 50 rounds (Figure~\ref{fig:The distribution and velocity over time}).

Interestingly, the cyclic pattern exists in all four stages. It is noteworthy that after 100 rounds, the N strategy is almost the most frequent strategy and the C strategy is always the most rare strategy, the cyclic pattern still exists. And in the third stage of T1, the pattern is so clear and similar to the pattern given by replicator dynamics ($\alpha$: $Coef.$=0.607, $p$=0.007; $\|\bar{v}^{o}\|$: $Coef.$=0.109, $p$=0.000).

The state change is like the whirlwind. The frequencies of strategies change quickly while the distribution of states in the state space moves spirally and slowly. From the first 50 rounds to the last 50 rounds, the overall distribution of social states changes in the state space and, from cooperation as the frequent strategy to defection and from defection to nonparticipation, forms a clear rotation path in a long run. The rotation path of state distribution seems to be ending at the N strategy, fewer remaining in the cycles. So, as time goes on, the distribution rotate from C to D, and from D to N slowly, meanwhile, the cycles become clearer from the second 50 rounds and persist over 200 rounds though there are more noise in the last 50 rounds.
\subsection{The individual's strategy cycles}

The results of cyclic pattern of social state change are shown above. However, it is still not clear that does the individual use the strategies cyclicly as well? Supposedly, the individual's choice to select one strategy will increase the frequency of that strategy in the population, and therefore will push the individual himself to turn to the more profitable strategy at that state. The individual will find the defection is more profitable when he chooses C, then he might turn to  choose D, and he might turn to choose N when he chooses D after finding other people choose D as well. We use the time series data of individual decision to detect the order of strategies. The full set loop is (C-C, D-D, N-N, C-D-C, C-N-C, D-C-D, D-N-D, N-C-N, N-D-N, C-D-N-C, C-N-D-C, D-N-C-D, D-C-N-D, N-C-D-N, and N-D-C-N).  Among them, C-D-N-C, C-N-D-C, D-N-C-D, D-C-N-D, N-C-D-N, and N-D-C-N form the real strategy cycle, in which, C-D-N-C, D-N-C-D, and N-C-D-N are anticlockwise and C-N-D-C, D-C-N-D and N-D-C-N are clockwise.

We can trace the individual's behaviors round by round in T1. The number of Rock-Paper-Scissors type loops (anticlockwise) is 180, while the number of anti-Rock-Paper-Scissors type loops (clockwise) is 72, the net number of Rock-Paper-Scissors type loops of individual's strategy is 108. Table~\ref{tab:individual cycle} exhibits the number of these cycles and the ratio of each type of the cycle. These results implicate that the individual's strategies present the characteristic of the cycle, and the direction is consistent with the cycle of social state change.

\begin{table}[htbp2]
\centering
\begin{threeparttable}
\small
\caption{\label{tab:individual cycle} The Individual's Strategy Cycles}
\begin{tabular}{cc|cccc}
  \hline
   \hline
Cycle Type	    &	        &Frequency	&	  &Ratio    &		\\
   \hline
Anticlockwise	&	        &180	    &	  &0.714	&    	\\
	            &C-D-N-C	&	        &95	  &	        &0.377		\\
	            &D-N-C-D	&	        &77	  &	        &0.306		\\
	             &N-C-D-N	&	        &8	  &	        &0.032		\\
   \hline
Clockwise	    &	        &72	        &	  &0.286	&    	\\
	            &C-N-D-C	&	        &34	  &	        &0.135		\\
	            &D-C-N-C	&	        &31	  &	        &0.123		\\
	            &N-D-C-N	&	        &7	  &	        &0.028		\\
   \hline
Total	        &	        &252	    &	  &1	        \\
   \hline
   \hline
Net Anticlockwise&	        &108	    &	  &0.429	&    	\\
	            &C-N-D-C	&	       &61	    &	    &0.242		\\
	            &D-C-N-C	&	       &46	    &	    &0.183		\\
	            &N-D-C-N	&	       &1	    &	    &0.004		\\
 \hline
  \hline
  \end{tabular}
     \end{threeparttable}
  \end{table}

So, the Rock-Paper-Scissors type of cycle exists not only at the aggregate social level but also at the individual level.

\section{Discussion and conclusion}

\subsection{Cyclic dominance and cooperation sustaining}

The mechanism by which nonparticipant can sustain cooperation is the Rock-Paper-Scissors type of cyclic dominance of the three strategies. That is, if the cycle is existent, the cooperation will always sustain. Otherwise, the cooperation cannot sustain stably in a long run even it may emerge occasionally. From the results above, it seems that, if the cycle is perfect, the cooperation is stable. For example, the cyclic pattern in the state space in the third stage in T1 is more perfect than other stages, while the cooperation in this stage is more stable. And the cyclic pattern of T1 is more perfect than that of T2, while the cooperation in T1 sustains better than T2.

The parameter used here is not the best for forming cyclic strategy evolution and therefore maintaining cooperation. The parameter used here, as mentioned in experimental design, is just the parameter that Sasaki et al.~\cite{sasaki2012take} used in their theoretical model which forecasts the existence of cycle as well as the outcome of strategy N. This parameter gives us more opportunities to further investigate the cooperation condition based on the Rock-Paper-Scissors type of cyclic dominance of the three strategies. For example, if we raise $r$, the cooperation might be more invited after N as Christoph Hauert et al~\cite{hauert2002volunteering} has pointed out. Furthermore, to introduce institutional incentive, i.e. reward or punishment as Sasaki et al.~\cite{sasaki2012take} has studied, or the underlying population structure as Szab{\'o} and Hauert~\cite{szabo2002evolutionary} studied, this parameter presents the marginal condition for forming strategy cycle. If the empirical cycle can be detected in this condition, then the existence of strategy cycle in the better condition is in prospect, and the diverse model based on these dynamics equations can be studied experimentally.

\subsection{Group size, population size and cooperation}

From the perspective of the individual, it is easy to understand why state change from C to D, and from D to N. However, it is difficult to understand why state will change from N to C. Indeed, it is really happened, but what is the real reason?

One possible explanation for the change from N to C is that when N strategy is more frequently used, the group size becomes small. The smaller the group is, the easier the reputation establishes. Is cooperation more favorable in smaller group? Table~\ref{tab:cooperation frequent in smaller group} exhibits the different cooperation frequencies and ratios under five groups sizes.

\begin{table}[htbp2]
\centering
\begin{threeparttable}
\small
\caption{\label{tab:cooperation frequent in smaller group}  The cooperation under five conditions}
\begin{tabular}{c|rrr|rrr}
  \hline
   \hline
	&	&T1	&	&	&T2	&	\\
\hline
Given N	&$n_{c}$	&$n_{d}$	&${n_{c}}/{n_{c+d}}$	&$n_{c}$	&$n_{d}$	 &${n_{c}}/{n_{c+d}}$	\\
\hline
0	&1439	&1546	&0.482	&150	&600	&0.200	\\
1	&692	&1336	&0.341	&190	&834	&0.186	\\
2	&341	&901	&0.275	&137	&772	&0.151	\\
3	&150	&498	&0.231	&75	    &433	&0.148	\\
4	&62	    &344	&0.153	&25	    &156	&0.138	\\
  \hline
 \hline
  \end{tabular}
     \end{threeparttable}
  \end{table}

In Table~\ref{tab:cooperation frequent in smaller group}, the first column is the condition given by the number of nonparticipators. If the number equals to 5, that means no one can choose C or D, so the range is from 0 to 4. The second column is the total number of people choosing C and the third column is the total number of people choosing D, and the fourth column is the ratio of the number of of people choosing C over all participators in T1. For example, given the condition that the number of nonparticipator is 0 for T1, the number of people choosing C is 1439, and the number of people choosing D is 1546, the ratio $C/(C+D)$ is 0.482. And the T2 is organized  in the same way. We find no evidence to support the point of view that the cooperation is more easy to establish in smaller group. On contrary, it seems that the cooperation is more frequent in large group.

Actually, the group size here is changing all over the time, whereas the population size is fixed. The population size of T1 is 5 and of T2 is 13. As mentioned above (see Table~\ref{tab:strategyusesummarize}, the cooperation rate and conditional cooperation rate of T1 are higher than those of T2. It indicatess if the population is large, the cooperation is low. This results are similar to John Duffy's work~\cite{duffy2009cooperative}. The cooperation established in a small group depends on the direct or indirect reciprocation. And this needs the glutinousness condition that the probability of another round is sufficiently high~\cite{axelrod1981evolution,wedekind2000cooperation}. Small is the necessary condition, but not the sufficient condition. If people cannot believe the person in the group, the round will still stay in the group in next round, the reputation can hardly be established. So, the question why state changes from N to C remains for further investigate.

\subsection{The bridge between evolutionary theory and experimental research}

Theoretically, the cyclic dominance of the three strategies is provided by evolutionary dynamics such as replicator dynamics. Although, mathematical exploration of evolutionary models has been a hot topic of research recently (see, e.g., Weibull, 1995), but it could not catch held of the experimenters. Camerer has pointed out that \textit{evolutionary models apply best to animals with genetically heritable strategies, or to human cultural evolution (Boyd and Richerson, 1985), neither of which explains rapid individual learning in the lab} (see Camerer~\cite{Camerer2003} p268). Huyck also emphasized: \textit{A lesson from the experiment is that one should discount models that predict deterministic cycles} (p148 in \cite{Huyck1999}). Since \textit{the subjects don't exhibit the kind of correlated behavior predicted by the dynamic} (p139 in \cite{Huyck1999}). On the other hand, Borgers argued that, in an appropriately constructed continuous time limit, the learning process' actual movement will not differ from its expected movement~\cite{Borgers1997}.

In this article, we provided the empirical cyclic dynamic pattern in full state space experimentally. And the result is very similar to the result given by replicator dynamics. It seems possible to establish the bridge between evolutionary theory and experimental research. If we use the new statistic measurement, for example the velocity, and get sufficient data, we can find the way to investigate the empirical dynamic pattern and can compare it with the evolutionary dynamics.

\subsection{Conclusion}

In this article, we have investigated the mechanism experimentally in the optional public goods game, and have shown, first time, the cyclic strategy pattern in full state space. It is also demonstrated that the strategies of cooperation, defection and nonparticipant form a Rock-Paper-Scissors type cycle, and the cycle of three strategies are persistent over 200 rounds. This cycle is very similar to the cycle given by evolutionary dynamics, for example, replicator dynamics. And the Rock-Paper-Scissors type cycle is not only existent at social level, but also at the individual level, that means the individual's strategies sequence of anticlockwise cycle of N-C-D-N exceeds it's adverse cycle. Meanwhile, the distribution of social states changes in the state space and, from cooperation as the most frequent strategy to defection and from defection to nonparticipant, forms a clear rotation path in a long run. Our results seem to implicate that the evolutionary dynamics has ability to capture the real dynamics not only on biosphere, but also on human society. This investigation provides the base structure to study the mechanism to sustain cooperation by evolutionary dynamics.

\textbf{Acknowledgment:} The work was supported by a grant of philosophy and social sciences planning project of Zhejiang province (13NDJC095YB), the 985 Project at Zhejiang University and by SKLTP of ITPCAS (No. Y3KF261CJ1) as well as the Discipline Construction Funds from College of Public Administration of Zhejiang Gongshang University. I thank Zhijian Wang for helpful discussion and also thank Zunfeng Wang for technical assistance.
\\
%\textbf{Reference}

\bibliographystyle{model1a-num-names}
%\bibliography{sasaki20130603}

%http://www.dictionaryofeconomics.com


%% References


%\bibliographystyle{unstr}






% \end{thebibliography}

\end{document}